\documentclass[twocolumn]{aastex701}

\usepackage{xcolor}
\usepackage{amsmath}
\usepackage{array}
\usepackage{stackengine}

\begin{document}

\title{The GRB Intrinsic Duration Distribution: Progenitor Insights Across Cosmic Time}

%The GRB Intrinsic Duration Distribution Suggests Progenitor Differences at Low and High Redshifts

\author[orcid=0000-0003-1707-7998]{Nicole M. Lloyd-Ronning}
\affiliation{Computational Physics and Methods Group, Los Alamos National Lab, 87545, US}
\affiliation{Center for Theoretical Astrophysics, Los Alamos National Lab, 87545, US}
\email[show]{lloyd-ronning@lanl.gov}  

\author[orcid=0000-0003-4271-3941]{Omer Bromberg} 
\affiliation{The Raymond and Beverly Sackler School of Physics and Astronomy, Tel Aviv University, Tel Aviv 69978, Israel}
\email{omer@wise.tau.ac.il}

\author[0000-0002-7964-5420]{Tsvi Piran}
\affiliation{Racah Institute of Physics, The Hebrew University, Jerusalem 91904, Israel}
\email{tsvi.piran@mail.huji.ac.il}

%\author[0000-0000-0000-0003,sname=Asia,gname=Mountain]{Asia Mountain}
%\altaffiliation{Astrosat Post-Doctoral Fellow}
%\affiliation{Tata Institute of Fundamental Research, Department of Astronomy}
%\email{fakeemail5@google.com}

%THIS HAS THE SHORTER ABSTRACT AND THE FIGURES AREN'T CALLED PROPERLY, BUT ARE ONCE IT'S UPLOADED TO APJ SITE
\begin{abstract}
%APJ Abstract Shorter:
We present the distribution of the {\em intrinsic} duration of gamma-ray burst emission.  This expands upon the analysis of \cite{Brom12} and \cite{Brom13} who showed evidence for collapsar progenitors based on the presence of a plateau in the distribution of $T_{90}$, the duration over which 90 \% of the prompt emission is observed.  We confirm the presence of this plateau in the distribution of duration corrected for cosmological time dilation, but shifted to smaller timescales by a factor of $1/(1+z_{\rm av}) \sim 1/3$, where $z_{\rm av}$ is the average GRB redshift.  More significantly, we show this plateau is only present in the sample of GRBs with redshifts greater than $(1+z) \sim 2$. This result aligns with suggestions that the low redshift population of GRBs has a significant contribution from non-collapsar progenitors (while the high redshift sample is dominated by collapsars). We also show the difference in this distribution between spectrally hard and soft GRBs, confirming that a plateau is only present for the soft subset of GRBs. However, when we separate the soft GRBs into low and high redshift subsets, we find that only the high redshift soft GRBs show evidence of a plateau, while the low-redshift soft GRBs do not.  This suggests {\em there exists a significant subset of spectrally soft non-collapsar progenitors at low redshift}.  Finally, we use the end time of the plateau to constrain the GRB progenitor density profile and radius, and show the maximum size of a collapsar is a few tenths of a solar radius.

\end{abstract}

%\keywords{\uat{Gamma-ray Bursts}{629} }

\section{Introduction}
Despite many thousands of observations of gamma-ray bursts (GRBs), the most luminous events in our universe, there still remain fundamental questions about the nature of their progenitor systems.  We know that many long (duration lasting more than about 2 seconds), spectrally soft GRBs likely result from collapse of massive stars, based on the presence of coincident supernovae in their spectra and light curves \citep{Gal98,Hjorth03, WB06, HB12} as well as their locations in star forming regions in their host galaxies \citep{Bloom02,Ly17}. We have similar circumstantial evidence for short (gamma-ray duration lasting less than about 2 seconds), spectrally hard GRBs associated with compact object mergers, in particular from their locations on the outskirts of their host galaxies, expected from the system's proper motion during the time between formation and merger, first suggested by \cite{NPP92} and later confirmed observationally \citep[e.g.][]{Fong2010, Fong2013}. And we have firm evidence that at least one short GRB, GRB170817, is associated with the merger of two neutron stars based on the coincident gravitational waves detected alongside this GRB\footnote{We note, however, this may not be a typical short GRB due to its low luminosity, soft spectrum, and unusual time profile \citep[e.g.][]{Kasliwal2017}.} \citep{Ab17}.\\

However, the traditional assumption that short, hard GRBs are associated with binary compact object mergers, and that long, soft GRBs result from a massive star collapse has been upended by a number of recent observations and theoretical studies, including evidence of a kilonovae (typically associated with binary neutrons star mergers) in long duration GRBs \citep[][although see \cite{Ris25}]{Rast22, Troja22, Yang22, Zhang2022, Yang24, Levan24}, a supernova associated with a short GRB \citep{Ross22}, and evidence that compact object mergers appear to be the reason behind the uptick in the long GRB rate density at low redshift \citep{Pet24, LR24, Chen24}. Additionally, \cite{Dimple22} and \cite{Dimple24} showed that some short GRBs at higher redshifts exhibit characteristics of collapsar progenitors.  There are also suggestions that certain sub-classes of GRBs potentially originate from distinct progenitors, including low-luminosity GRBs \citep{Sod06,Guet07}, radio bright and radio dark GRBs \citep{chakraborty2022,LR22}, short GRBs with extended emission \citep{Metz08,Kan15}, and one GRB that repeated multiple times over a 24 hour period \citep{Lev25}. \\

Furthermore, and importantly, some GRBs {\em appear} as long GRBs, but in fact have very short prompt duration when corrected for cosmological time dilation.  Two excellent examples are the very high redshift GRB120923A and GRB090429B. The former, with a measured prompt duration of $27.2$ s is at a redshift of $(1+z) = 8.8$ \citep{Tan18}, bringing its intrinsic duration down to just $3$ seconds.  Meanwhile, GRB090429B is measured at a redshift of $(1+z) = 10.4$ \citep{Cuc11}, bringing its observed duration of $5.8$ seconds down to just $0.56$ seconds.  We discuss these bursts further below, but our point here is that any inferences based on classification by duration need to account for cosmological time dilation.\\

Recently, machine learning techniques have been developed to help better classify GRBs beyond just the short-hard/long-soft delineation \citep{Luo23, negro2024, Garc24, Zhu24,Chen24b, Esp25}.  These techniques often combine highly resolved timing and spectral information into multi-dimensional parameters space, in which ``nearest neighbors'' are computed to produce a type of map in which similar GRBs tend to cluster. However, using these methods to draw a straightforward connection to the nature of their progenitors is not necessarily clear.\\

 It is likely that many or all of the suggested GRB progenitor systems play a role in producing GRBs, and any distinct signatures of the progenitor are washed out in the cataclysmic event that produces the gamma-ray burst.  Nonetheless, efforts toward understanding the relative fraction of possible progenitors systems and how that fraction evolves over cosmic time is not only crucial to understanding GRB physics, but also the evolution of massive stars in general.\\

An important clue was offered by \cite{Brom12} and \cite{Brom13}, who showed that a plateau in the distribution of the observed GRB prompt duration, $dN/dT_{90}$, provides strong evidence for the presence of a collapsar progenitor\footnote{We note \cite{MP17} additionally showed evidence for a plateau in the distribution at around 0.4 seconds, interpreted as arising from the jet breaking out the ejecta around a compact binary merger.}.  The essential idea is that the observed duration $t_{\gamma}$ is the difference between the time over which the central engine is active, $t_{e}$, and the minimal activity time of the engine that allows the jet to break out of the stellar envelope, $t_{\rm th}$, what we call the ``threshold time''.\footnote{For a non-relativistic jet $t_{\rm th}$ is the actual breakout time from the stellar surface, however in the relativistic case $t_{\rm th}$ is shorter than the breakout time by a factor of $1-\beta$.} 
%\obn{I suggest to change the former sentence to: "The essential idea is that the observed duration $t_{\gamma}$ is the difference between the time over which the central engine is active, $t_{e}$ and the minimal activity time of the engine that allows the jet to break out of the stellar envelope, $t_{bo}$}\nlr{Done. Thank you.}
The probability that a GRB has a duration $t_{\gamma}$ is equivalent to the probability that the engine has a work time $t_{\rm th} + t_{\gamma}$; that is, $p_{\gamma}(t_{\gamma}) = p_{e}(t_{\rm th}+t_{\gamma})$. For observed durations less than a given threshold time, $p_{e}(t_{\rm th} + t_{\gamma}) \approx p_{e}(t_{\rm th}) + O(t_{\gamma}/t_{\rm th})$ is approximately constant.  In other words, $p_{\gamma}(t_{\gamma}) \approx p_{e}(t_{\rm th}) \approx {\rm constant}$, for $t_{\gamma} < t_{\rm th}$. Provided the probability distribution of $t_{e}$ is a relatively smooth function, this necessarily leads to a flat distribution of duration for any engine duration times less than the threshold time of the star \citep[][show that this claim holds even for a distribution of threshold times]{Brom12}.  In other words, under these conditions $t_{\rm th}$ will be the time where the plateau will end in the $p(t)$ distribution. 
%\obn{If you accept my previous commen t, than the previous sentence is obsolite. Also, we need to decide if we're using $t_{bo}$ ot $t_{th}$.}\nlr{Yes, let's use $t_{\rm th}$, thanks.} 

The probability distribution $p(t)$ is equivalent to $dN/dt$, the differential number of observed GRBs per unit duration $t$, not per log ($t$) as is normally shown in the literature\footnote{in which case the plateau becomes a linear increase.} \citep[e.g.][]{Kouv93}. As discussed in \cite{Brom12} and in \S 4.3 below, this threshold timescale can provide constraints on the progenitor stellar radius and density profile, which -- in turn -- provide further constraints on the type of stellar system that can produce a GRB.\\
%\footnote{It can be especially constraining if collapsar progenitors have a fairly narrow distribution of threshold times.  For more broadly distributed threshold times, it may only help constrain the minimum breakout time of a GRB collapsar progenitor.} 

In this paper, we perform a similar analysis to \cite{Brom12} and \cite{Brom13}, but for the so-called {\em intrinsic duration} distribution, the observed duration corrected for cosmological time dilation.  We show that, as expected, this lowers the value of the timescale at which the end (or edge, at the high duration side) of the plateau is evident in the distribution.
%\footnote{What we call the ``edge'' is where the plateau begins to appear in going from high to low duration, or left to right on our $dN/dT$ plots in Figures 1 - 4.}
%\obn{I suggest to change to : "this lowers the duration of the plateau"}\nlr{Ok, looking at the comments below it is clear we need to discuss this wording.  I just need to add a definition of what I mean by ``onset'' or begins.  See my response to your next comment below.}; 
When analyzing our full sample of about 500 GRBs with measured redshifts, the end of the plateau is around a few seconds (compared to $\gtrsim 10$ seconds seen in the observed $T_{90}$ distribution)

More revealingly, we show that when our sample is divided into ``low'' and ``high'' redshift subsets (those with $(1+z) \lesssim 2$ and those with $(1+z) \gtrsim 2$), the plateau only appears in the high redshift sample.  The fact that the plateau is not present (or only weakly so) in the low redshift sample supports the conjecture that the population of low redshift GRBs contains a significant fraction of non-collapsar systems, likely compact object mergers, while the sample of high redshift GRBs is composed predominantly of collapsars. When our sample is divided into spectrally ``hard'' and ``soft'' subsets, the hardest GRBs do not show evidence of a plateau, while the soft GRBs do; at first glance, this appears to support previous studies claiming spectral hardness is a good indicator of progenitor type.  However, when we further divide the spectrally hard and soft samples into low and high redshift subsets, the hard GRB subset does not show evidence of a plateau at either low or high redshifts. The soft sample, on the other hand, shows a clear plateau {\em only in the high redshift subset}, suggesting the existence of a significant population of spectrally soft non-collapsar progenitors at low redshift.  This discovery challenges the traditional classification of collapars and non-collapsars based on the spectral shape or the hardness ratio \citep{Kouv93}. \\

Our paper is organized as follows: In \S 2, we present our data sample, methods, and discuss potential selection effects and sample completeness.  In \S 3, we present our results: the distribution of the prompt intrinsic (corrected for cosmological time dilation) duration for our whole sample; the distribution of intrinsic duration for GRBs separated into low and high redshift sub-samples; the distribution of intrinsic duration for GRBs separated into spectrally hard and soft GRBs, with the latter also divided into low and high redshift sub-sets.  In \S 4, we present the physical implications of our results. We discuss the ramifications of the presence of a plateau in the higher redshift samples, the lack of a plateau in the lower redshift samples, and how this aligns with studies that have suggested the uptick in the low redshift GRB rate density may be hinting at compact object merger progenitors for these GRBs (regardless of duration and spectral hardness). We also discuss how the end time of the plateau can constrain properties of the progenitor star and provide limits on the collapsar radius, given the average observed properties of the jets in our sample.   Finally, in \S 5, we present our summary and conclusions.  Our Appendix provides supplementary data and analysis.

\section{Data and Methods}

Our data are taken from \cite{Wang20}, who have compiled publicly available observations of $6289$ gamma-ray bursts from 1991 to 2016.   We searched this dataset for those GRBs with measured redshifts, the vast majority of which are GRBs observed by the {\em Neil Gehrels Swift} Observatory (492 {\em Swift} GRBs with redshifts out of the full sample of 567 GRBs with redshift values).  We then take the measured $T_{90}$ values (the timescale over which 90\% of the prompt emission is observed) and correct it for cosmological time dilation to get the intrinsic duration, $T_{int}$ in the (cosmological) rest frame of the progenitor: $T_{int} = T_{90}/(1+z)$.  From this, we construct the duration distribution $dN/dT_{int}$.

\begin{table*}[ht]
\centering
\caption{Parameters from equation~\ref{eq:func}, fit to the different $dN/dT$ distributions.  We do not list the values of $\beta$ in the table below as all the fits returned $\beta = 0.$
%{\bf Note that I did not constrain any of the parameters in any way (e.g. to remain positive), which is why in some cases the $A_{NC}$ and $\sigma$ coefficients are negative (but still provide a reasonable fit).} 
Note for the low redshift and hard sub-samples, the value of $T_{B}$ is relatively low but has an error bar many times greater than the fit value, indicating there is no clear plateau in these data.}
\begin{tabular}{lcccccccc}
\hline
\textbf{Dataset} & \textbf{$A_{NC}$} & $\boldsymbol{\mu}$ & $\boldsymbol{\sigma}$ & \textbf{$A_{C}$} & $\boldsymbol{\alpha}$  & \textbf{$T_{B}$ (s) } & $\boldsymbol{\chi^2}$/dof \\
\hline
\hline
$T_{90}$  & -28.9 $\pm$ 13.5 & 2.3 $\pm$ 1.5 & -4.3 $\pm$ 2.7 & 3.1 $\pm$ 1.0 & -0.4 $\pm$ 0.1 &  {\bf 8.2 $\pm$ 5.1} & 1.01 \\
$T_{int}$ &  3.1 $\pm$ 3.1 & -0.5 $\pm$ 9.3 & 6.2 $\pm$ 10.9 & 1.0 $\pm$ 0.3 & -0.5 $\pm$ 0.1 &  {\bf 2.5 $\pm$ 1.1} & 0.87\\ 
\hline \hline
$T_{int}$, Low z  & 5.7 $\pm$ 3.0 & 1.4 $\pm$ 0.8 & 2.9 $\pm$ 0.8 & 0.1 $\pm$ 0.4 & -0.3 $\pm$ 0.5 &  {\bf 4.1 $\pm$ 19.2} & 0.33 \\
$T_{int}$, High z & 50.3 $\pm$ 19.2 & 2.4 $\pm$ 0.3 & 1.8 $\pm$ 0.3 & 1.3 $\pm$ 0.8 & -0.5 $\pm$ 1.1 & {\bf 15.8 $\pm$ 6.6}  &  0.67 \\ 
\hline \hline
$T_{int}$, Hard &  -156.0 $\pm$ 33969.8 & -13.0 $\pm$ 4318.2 & -29.7 $\pm$ 5644.2 & -0.1 $\pm$ 0.7 & -1.0 $\pm$ 23.5 &  {\bf 8.0 $\pm$ 126.6} & 0.97\\
$T_{int}$, Soft &  92.1 $\pm$ 42.5 & 2.1 $\pm$ 0.5 & 1.8 $\pm$ 0.3 & 2.4 $\pm$ 1.5 & -0.6 $\pm$ 0.9 &  {\bf 18.0 $\pm$ 9.7} & 0.56   \\
\hline \hline
$T_{int}$, Soft, Low z &  42.3 $\pm$ 68.0 & 1.6 $\pm$ 2.7 & 2.6 $\pm$ 2.2 & 1.5 $\pm$ 5.9 & -0.3 $\pm$ 1.1 &  {\bf 2.4 $\pm$ 11.5} & 1.70  \\
$T_{int}$, Soft, High z &  5.3 $\pm$ 2.9 & 1.3 $\pm$ 0.1 & 0.3 $\pm$ 0.2 & 1.3 $\pm$ 0.3 & -0.3 $\pm$ 0.7 &  {\bf 13.7 $\pm$ 8.0}   &  0.82\\
\hline
\end{tabular}
\label{tab:funcfit}
\end{table*}

\subsection{Selection Effects and Sample Completeness}

It is essential to examine the biases present in the data and how they can affect our analysis.  There are several issues to consider. \cite{Brom13} showed that the plateau is detector dependent. Indeed, observed duration is detector dependent as instruments not only have different flux sensitivity limits, but their bandpass sensitivities will also affect measured duration.  For example, those detectors sensitive to photons over an energy range that aligns with the average peak of the GRB $\nu F_{\nu}$ distribution (i.e. around a few hundred keV) will tend to have longer $T_{90}$'s. We perform our analysis on {\em only} {\em Swift} GRBs to account for this effect.\\

There is, of course, always a Malmquist-like bias present in any observation (i.e. objects farther away appear fainter).  However, the broad GRB luminosity function (the fact that GRBs are not standard candles) largely mitigates this bias. In terms of analyzing the duration distribution of GRBs, this effect play a role through an artificial shortening of GRB duration with redshift, also known as the ``tip-of-the-iceberg'' effect \citep{Bonn97}.  The severity of this effect, where GRBs at high redshift may appear to have shorter duration because some of their flux is shifted below the detector threshold limit, has been discussed in several studies, including \cite{NP02, NP02b, Pir04, Littlejohns13, Moss22, LR23, Moss26}.  

\cite{NP02, NP02b} found that the typical long GRB duration was dominated not by the combined length of the pulses but by the combined lengths of intervals between pulses. They also found that second highest peak in a long GRB is not significantly smaller than the highest one, so that if we see the first peak we are likely to see the second. As such, they concluded the tip-of-the-iceberg effect is not causing an outsized bias in the overall duration distribution. Additionally, \cite{Littlejohns13} found that they could not demonstrate that the high redshift duration sample was simply a low redshift duration sample that was shifted to higher redshifts (i.e. that the high redshift duration distribution was strongly influence by this effect).  Finally, \cite{LR23} corrected for this bias using non-parametric methods of \cite{LB71} and \cite{EP92}, and found that this bias does not have a strong effect on the trends seen in the duration distribution over cosmic redshift. They suggested that, again, because GRBs are not standard candles this effect is weakened and its influence on the behavior of the overall duration distribution is relatively small.  Recent works by \cite{Moss22} and \cite {Moss26}, however, claim that $T_{90}$ is indeed sometimes significantly underestimated at high redshift.\footnote{We note they impose some spectral constraints on their sample that may exacerbate the effect in their population simulations.}

Even assuming this effect does play a significant role in the observed duration, such a result tends to strengthen our arguments to some extent.  We show below that at redshifts larger than $\sim 2$, a plateau in the duration distribution extends to intrinsic prompt durations less than two seconds. We conclude from this that, at these redshifts, one cannot rely on the naive classic two second delimiter between short and long GRBs, because GRBs with intrinsic short durations at high redshifts are more likely to be of collapsar origin. \\

A potentially more important selection effect is one against detecting classically short GRBs ($T_{90} \lesssim 2s$) at high redshifts. The issue here is not so much a problem of short GRBs having lower prompt gamma-ray luminosity (and therefore being harder to trigger on or detect at high redshift).  In fact, several studies have shown short GRBs luminosities are comparable to long GRBs, especially in terms of the break in the luminosity function \citep{WP15,Ghir15, Dav15, Ghir16}, although see \cite{WP15,Dai21}, who show that the short GRB luminosity function falls more steeply above the break than that of long GRBs, which can cause a bias against detecting high redshift short GRBs.  

However, perhaps the more important problem lies in their lack of afterglow detection. 
Many short GRBs are found in the outskirts of their host galaxies \citep{Fong2010, Fong2013, Cast25}, in environments of lower circumburst density.  Additionally, short GRBs have on average smaller isotropic-equivalent energies $E_{iso}$ than long GRBs. Both of these factors make their afterglows harder to detect and therefore a redshift determination untenable.  Because we are examining only the sample of GRBs with redshift measurements, this bias may certainly affect our sample. Interestingly, for the few short GRBs with opening angle measurements, the beaming corrected energy is not much smaller than that of long GRBs; 
furthermore, there are quite a few short GRBs at large offsets from their host galaxies that still show observable afterglow emission \citep{Fong2013}. 
Therefore, the extent of an observational bias against detecting high redshift short GRBs is not entirely clear.  We revisit this issue -- and how it can affect the interpretation of our results -- further below in \S 4. \\

Finally, we also separate our sample by hardness classes, as in \cite{Brom13}. Definitions of spectral hardness vary depending on the instrument detecting the GRB (and its detector response as a function of energy).    Because our sample is comprised of GRBs detected by {\em Swift}, which is sensitive to relatively low gamma-ray energies compared to other GRB detectors, we use the ``steepness'' of the photon spectral index according to either a single power-law or cutoff power-law fit (discussed in more detail in the next section).  However, it should be noted that spectral hardness can be inherently difficult to characterize.  For example, because {\em Swift} is only sensitive to energies of  up to 200 keV, whereas the peak energy of a GRB spectrum often falls in the range of several hundreds of keV, there may exist GRBs with a relatively soft low energy photon index, but a very high peak energy or shallow high energy photon index which would indicate the burst is much harder than it appears in the {\em Swift} bandpass.   \\
%There is also a sample completeness issue here (not that many {\em Swift} bursts with redshifts have the single power law fits used in Bromberg et al. 2013). 

With these potential biases in mind, we proceed with our analysis of the $dN/dT$ distribution (and discuss further below where they may play a role in shaping our results), again focusing on the sample of {\em Swift} GRBs with redshifts. 
%We acknowledge, however, the inherent spectral biases that arise because of its lower bandpass sensitivity, and keep that in mind in interpreting our analysis below.   

\subsection{Quantifying the Timescale of the Plateau}

To quantify the time when the plateau ends we fit the duration distribution of our samples with a combination of two analytic functions for the distribution of Collapsars and non-Collapsars, taken from \citet[][Eq. 1]{Brom13} and shown below. The non-collapsar (NC) component modeled by a log-normal distribution, while the collapsar (C) component is modeled by a constant below some characteristic time $T_{int} \leq T_B$ and a power-law plus exponential cutoff above this time $T_{int} > T_B$:

\begin{multline}
\frac{dN}{dT_{int}} = A_{\text{NC}} \frac{1}{T_{int}\sigma\sqrt{2\pi}} e^{-\frac{(\ln T_{int}-\mu)^2}{2\sigma^2}} \\
+ A_C \begin{cases}
1 & T_{int} \leq T_B \\
\left(\frac{T_{int}}{T_B}\right)^\alpha e^{-\beta(T_{int}-T_B)} & T_{int} > T_B,
\end{cases}
\label{eq:func}
\end{multline}
 We perform a weighted nonlinear least-squares fit using the equation above.  When the fit is reasonable (in the sense of a small reduced $\chi^{2}$, and that each component of the function is a good description of the total data), then $T_{B}$ roughly represents the time of end time of the plateau.

\section{Results}

\begin{figure}[ht]
\centering
  % \stackinset{<horizontal-spec>}{<h-offset>}{<vertical-spec>}{<v-offset>}{<inset-content>}{<base-content>}
 % \stackinset{r}{6.5cm}{t}{7.0cm}{
 %   \includegraphics[width=0.37\textwidth]{dNdTall_noline.png}
%  }{
    \includegraphics[width=\linewidth]{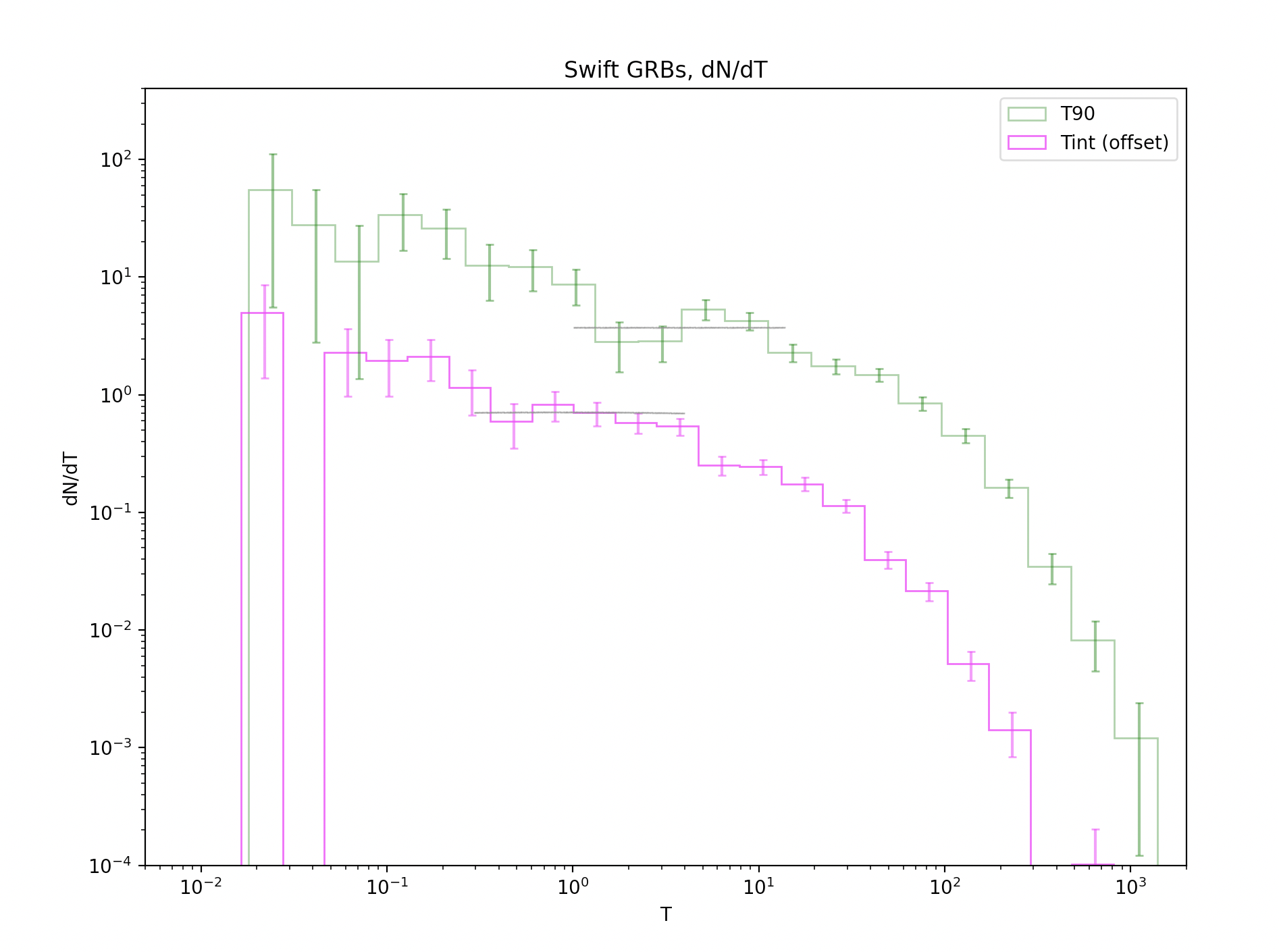}
%  }
  \caption{$dN/dT$ distribution of both $T_{90}$ (green line) and intrinsic duration $T_{int}$ (magenta line), for our sample of {\em Swift} GRBs with redshifts; the distributions are artificially offset by a constant along the y-axis for comparison purposes.   The right edge of the lines indicates the end of the plateau. The end of the plateau of the intrinsic duration distribution is shifted to shorter timescales by a factor of $\sim 1/3$. Table~\ref{tab:funcfit} gives the best-fit parameters of these distributions according to equation~\ref{eq:func}.}
  %Table~\ref{tab:funcfit} gives the best-fit parameters of these distributions according to equation~\ref{eq:func}, with the end of the plateau (defined by the parameter $T_{B}$) occurring at $\sim 2.5$ s for the $T_{int}$ distribution and $\sim 8$ s for the $T_{90}$ distribution.
  \label{fig:dNdT_total}
\end{figure}

\begin{figure}[ht]
 \centering
%  \stackinset{r}{6.3cm}{t}{7.7cm}{
%    \includegraphics[width=0.34\textwidth]{Figures/dNdT90redsepErr.png}
%  }{
%    \includegraphics[width=\textwidth]{Figures/dNdTredsepErr.png}
%  }
\includegraphics[width=\linewidth]{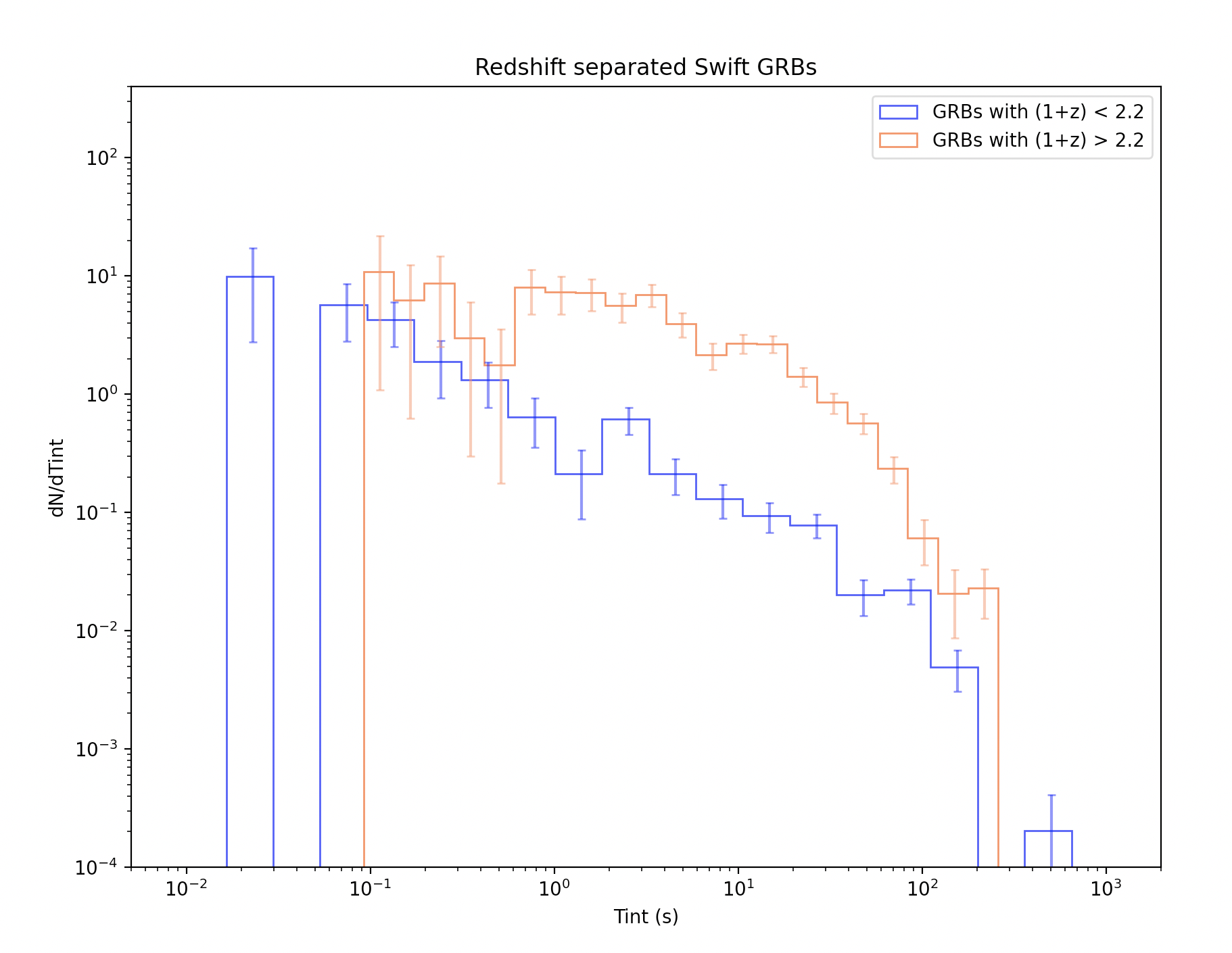}
\caption{$dN/dT$ distribution of intrinsic duration, $T_{int}$, broken into "low" redshift (blue lines) and "high" redshift (orange lines), with a redshift delimiter of $(1+z) = 2.2$.  There is a clear plateau in the high redshift sample, while no plateau present in the low redshift sample. The fits to these distributions, according to equation~\ref{eq:func}, are given in Table~\ref{tab:funcfit}.}
\label{fig:dNdT_zsep}
\end{figure}

\begin{figure}[h]
\begin{centering}
\includegraphics[width=\linewidth]{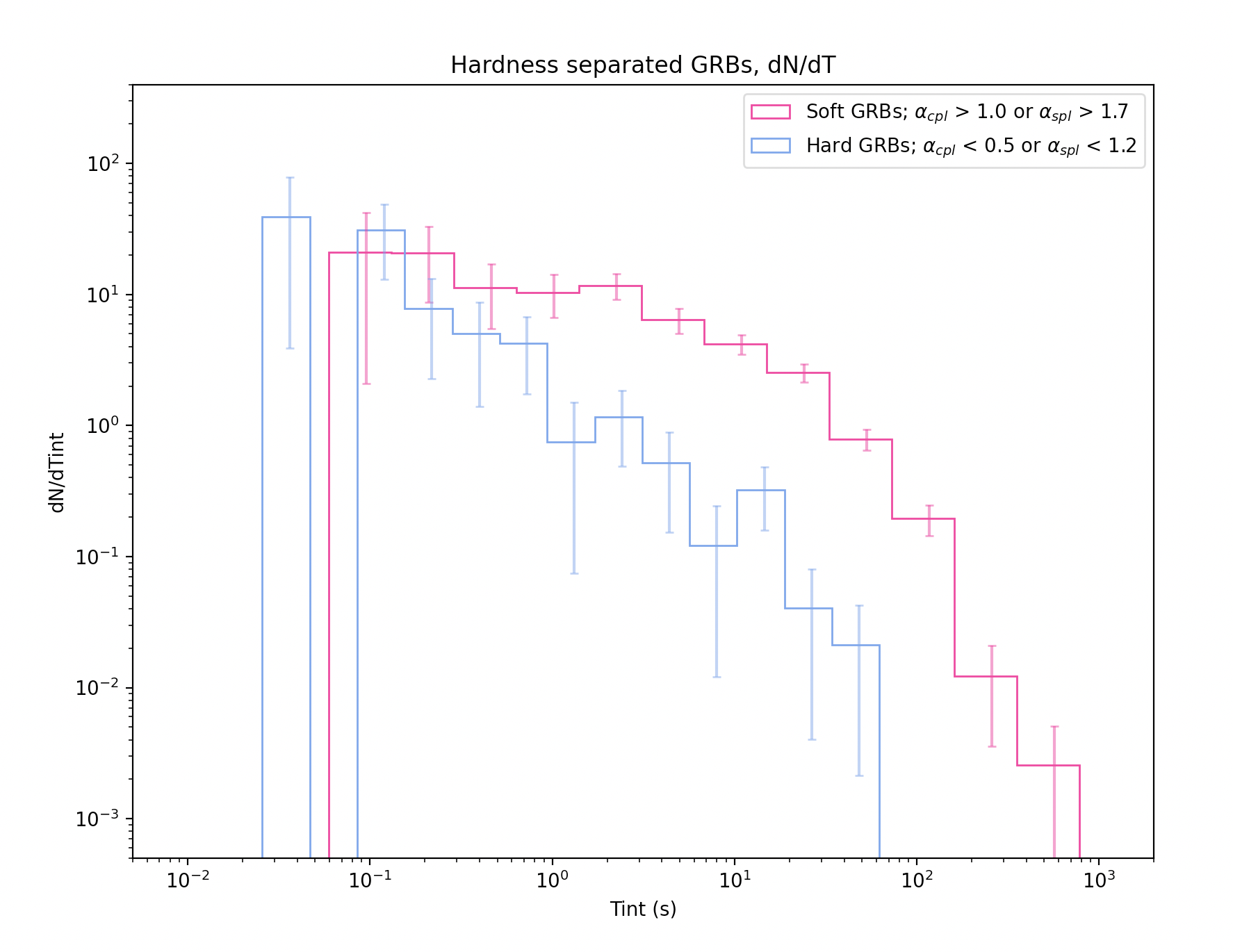}
\caption{$dN/dT$ distribution of intrinsic duration, $T_{int}$, for {\em Swift} GRBs with redshift, broken into ``hard'' and ``soft'' sub-samples based on the power-law index and spectral fit model. A plateau is present in the soft sample with an end at around a few seconds, while there is no plateau in the hard sample, confirming the results of \cite{Brom13}. The fits to these distributions, according to equation~\ref{eq:func}, are given in Table~\ref{tab:funcfit}.}
\label{fig:dNdThardness}
\end{centering}
\end{figure}

\begin{figure}[h]
\begin{centering}
\includegraphics[width=\linewidth, height=6.5cm]{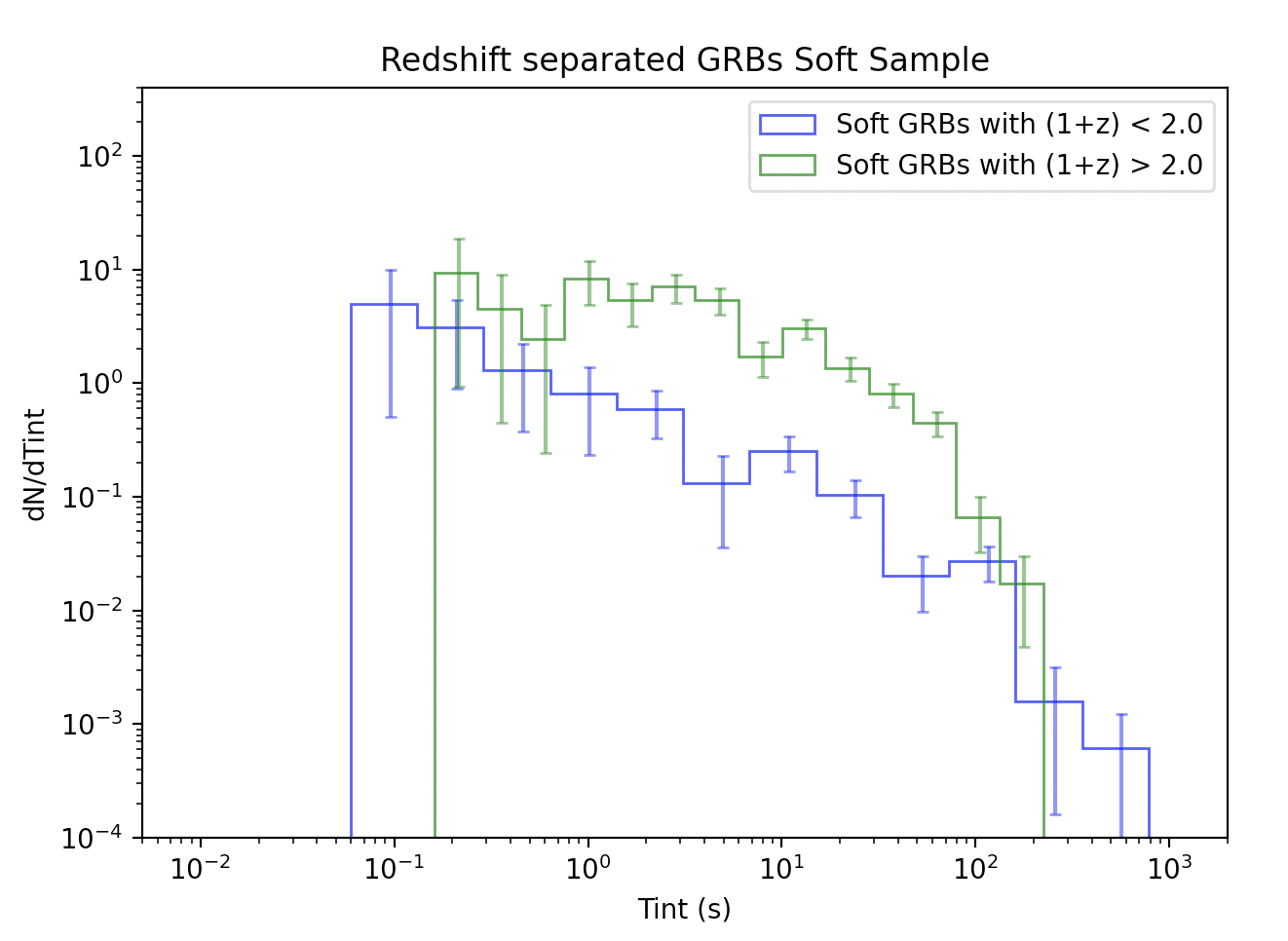}
\caption{$dN/dT$ distribution of intrinsic duration, $T_{int}$, for the {\em Swift} GRB spectrally soft sample broken into "low" redshift (blue lines) and "high" redshift (green lines), with a redshift delimiter of $(1+z) = 2.0$. Even within the soft sample we see a lack of a plateau in the low redshift GRBs, and a clear plateau in the high redshift GRBs, confirmed by our fits to equation~\ref{eq:func}, shown in Table~\ref{tab:funcfit}.}
\label{fig:dNdThardnessredsep}
\end{centering}
\end{figure}

Figures~\ref{fig:dNdT_total} - \ref{fig:dNdThardnessredsep} show our $dN/dT_{int}$ distributions, and Table~\ref{tab:funcfit} shows the best fits to these distributions according to equation~\ref{eq:func}.\\

Figure~\ref{fig:dNdT_total} shows the distribution of the intrinsic duration, $dN/dT_{int}$ (magenta), as well as observed duration, $dN/dT_{90}$ (green), for the {\em Swift} GRBs with measured redshifts, binned into equally spaced bins.  
The magenta and green histograms are artificially offset by one order of magnitude along the y-axis, for clarity of comparison, and we assume Poisson statistics to estimate the errors in each bin. 
The end of the plateau in the $T_{int}$ distribution appears at shorter durations than that of $T_{90}$, occurring around just a few seconds (compared to $\sim 10$'s of seconds in the $T_{90}$ duration).  The ratio between where the plateau ends occur (according to our fits shown in Table~\ref{tab:funcfit} and as evident by eye) is a factor of $\sim 1/3$, consistent with the correction due to cosmological time dilation for the average redshift of a GRB, $1/(1+z_{\rm av})$.
%\ob{Our best fitted models show plateau ending times of $T_B=2.5\pm1.1$ sec and $T_B=8.2\pm5.1 sec$ sec for the intrinsic and observed $T_{90}$ distributions respectively,  with a ratio of $T_{B,int}/T_{B,90}\sim1/3$ consistent with the correction due to cosmological time dilation for the average redshift of a GRB, $1/(1+z_{\rm av})$.}
We note that our $T_{90}$ distribution -- specifically where the end of the plateau occurs -- for those {\em Swift} GRBs with redshift measurements agrees with that of \cite{Brom12}, who show the $T_{90}$ distribution for all {\em Swift} GRBs both with and without redshift (see their Figure 1). Additionally, when we do a Kolmogorov-Smirnov test comparing our $T_{90}$ sample to the full Swift $T_{90}$ distribution, we find a p-value of 0.3, meaning we {\em cannot reject} the null hypothesis that these distributions are drawn from the same parent population. Stated more simply, our sample $T_{90}$ distribution (those GRBs with measured redshifts) and the entire Swift $T_{90}$ distribution are statistically equivalent.

\subsection{Redshift Separated Distributions}
Figure~\ref{fig:dNdT_zsep} shows the intrinsic duration distribution, $dN/dT_{int}$, for all of our {\em Swift} GRBs, divided into ``lower'' redshift (blue lines) and ``higher'' redshift (orange lines) sub-samples, with a delimiting value of $(1+z) = 2.2$.  The high redshift sample shows a clear plateau from $\sim 0.1$ s up to around $\sim 15$ s (confirmed by our fits shown in Table~\ref{tab:funcfit}), with no evidence of any uptick or ``non-collapsar'' increase at low durations. In contrast, and again confirmed by our fits using the function in equation~\ref{eq:func}, the low redshift distribution shows no evidence of a plateau. 
%\obn{The intrinsic distribution of the entire sample shows a plateaus that ends at $5$ sec and $T_B=2.5$ sec, while the intrinsic distribution of the high redshift sample shows a plateau that ends at 15 sec with best fitted $T_B=15$ sec. This is weird because the high redshift sample is included in the complete sample. I think that its mostly because the fit of the full sample underestimates $T_B$ and the error on $T_B$, while in the high $z$ sample is overestimates it. In any case I think we should address this apparent inconsistency in the text. We can discuss it.}

The distinction in the $dN/dT_{int}$ distributions between low and high redshift GRBs appears to show up most prominently when using a delimiting value of redshift between $(1+z) =2$ to $2.5$ (in the Appendix we show how this plot changes as a function of the redshift delimiter), suggesting that non-collapsar progenitors begin to play a more dominant role in producing GRBs below this redshift, while above this redshift, collapsars will dominate the progenitors of GRBs.  \\

The lack of short GRBs in the high redshift sample may be attributed to the observational biases discussed in \S 2. Nevertheless, there are strong physical reasons to expect a real deficit of high redshift short GRBs, due to long delay times between progenitor formation and compact-object merger, which result in a shift of a large fraction of these events to low redshift  \citep[on the order of a Hubble time for most of them;][]{Pir92, GP06,Nak06, Berg07, JL10, Cow12, HY13, WP15, Ana18, Bel18, Chr18, Broe22, Santo22, Zev22, Maoz24}. This means that most of the short GRBs (if the majority of this population comes from compact object mergers) are expected to occur at lower redshifts, and that this is not simply due to an observational selection effect.\footnote{Interestingly and aligned with our results, \cite{Dimple22} and \cite{Dimple24} showed that those short GRBs which are located at higher redshifts appear to have characteristics of collapsar progenitors.}\\

\subsection{Hardness Separated Distributions}
\cite{Brom13} was one of the first studies to suggest that there is a non-negligible fraction of long duration ($\gtrsim 10$ s) non-collapsar GRBs.  They showed this by separating GRBs according to their spectral hardness and demonstrating that the plateau (the telltale collapsar signature) is not present in the hardest subset of GRBs, regardless of duration, while it is present in the softest subset of GRBs. Hence, they suggest that spectral hardness might be a better indicator of progenitor class, compared to duration.  We perform a similar analysis here, for the spectrally separated duration distribution corrected for cosmological time dilation.  Because definitions of spectral hardness are detector dependent, we emphasize again that we are only analyzing the {\em Swift} GRBs in our sample. We have divided our bursts into hardness classes according to the steepness of its low energy spectral photon index, for two different models of the spectrum.  The first is a single power-law ({\em spl}):
\begin{equation}
    N(E) = A_{spl}E^{-\alpha_{spl}}
\end{equation}

\noindent and the second is a power-law with an exponential cutoff ({\em cpl}):
\begin{equation}
    N(E) = A_{cpl}E^{-\alpha_{cpl}}{\rm exp}^{-E/E_c}
\end{equation}
\noindent where $A$ is a normalization factor, and $E_{c}$ is the cutoff energy.
Each model has a different criterion for what constitutes spectral hardness.  In the Appendix, we show the spectral indices of as a function of $T_{90}$ for all GRBs for which these fits are available, and show where we have chosen to draw our cuts for what constitutes ``hard'' and ``soft'' for each model, where ``hard'' is defined as a value of $\alpha$ above the upper cutoff line and ``soft'' is defined as a value of $\alpha$ below the lower cutoff line.\\

Figure~\ref{fig:dNdThardness} shows our hard (light blue) and soft (pink) {\em Swift} GRBs.  Consistent with \cite{Brom13}, but -- again -- shown here for the intrinsic duration distribution, we see that the hardest GRBs show no evidence of a plateau, while soft GRBs show evidence of a plateau end at around a few seconds.  
%This is confirmed by the slope analysis of these distributions, shown in Figure~\ref{fig:dNdT_slopeanalys3}.
Figure~\ref{fig:dNdThardnessredsep} shows the $dN/dT_{int}$ distribution of our soft sample, but divided into low and high redshift groups.  As with the entire sample, we see that the lower redshift group does not show evidence of a plateau, confirmed by our fits shown in Table~\ref{tab:funcfit}.  This means that there exists a significant population of spectrally soft non-collapsar progenitors at low redshift. We note that when the spectrally hard GRBs are split into low and high redshift sub-samples, there is no evidence of a plateau in either distribution (at low or high redshift), suggesting spectrally hard GRBs are indeed dominated by non-collapsars across cosmic time. However, the number of GRBs in this sample is too small to put this statement on firm statistical footing.

\subsection{Summary of Results}
For the intrinsic duration distribution of our whole sample of {\em Swift} GRBs with redshift, the end of the plateau is (unsurprisingly) shifted to lower timescales by a factor of $\sim 1/(1+z_{av})$, where $z_{av}$ is the average redshift of the GRB population.  For the redshift separated distributions, the end time of the plateau in the {\em high redshift} sample appears at around $\sim 10$ to $15$ seconds, while there is no plateau apparent in the low redshift sample.  For the hardness-separated sub-samples, the plateau occurs at about $10$ seconds in the soft sample, while there is no apparent plateau in the spectrally hard sample. Finally, for the redshift-separated spectrally soft sample, the plateau remains in the high redshift sample (with an end time at around $\sim 10$ s), but there is no apparent plateau in the spectrally soft, low redshift sample.  All of these statements are quantified in Table~\ref{tab:funcfit} for the different sample subsets. We note that the fitting results do depend on the binning of the data and occasionally the function does not always characterize the data well.  For example, for the low redshift and hard sub-samples, the error bars on $T_{B}$ are many times greater than the value of $T_{B}$ itself.  This indicates that there is no firm $T_{B}$ value for these data -- in other words, there is no plateau in these distributions.\\

The primary takeaway from this analysis is that GRBs at redshifts $(1+z) \gtrsim 2$ are dominated by collapsar progenitors, while the GRB sample at redshifts $(1+z) \lesssim 2$ appear to be dominated by non-collapsar progenitors, regardless of spectral hardness.

\section{Physical Implications}

The fact that the plateau is evident in some of our sub-samples and not in others, combined with the plateau end time values when it {\em is} present has a number of implications. First, we note the disparity in the plateau end times between the different $T_{int}$ distributions (i.e. Figures~\ref{fig:dNdT_total} - ~\ref{fig:dNdThardness}), when the plateau is clearly present. This is likely a result of an imperfect fit of our analytic functions to the GRB duration distribution.  When the number of non-collapsars increases, it pushes $A_C$ to higher values (i.e. raises the height of the fitted plateau); as a result, the best-fit value of $T_B$ is reduced. In other words, the fit of $T_B$ better captures the true end time of the plateau in samples that are less contaminated by non-collapsars, such as the sample of high redshift GRBs (those GRBs with $(1+z) \gtrsim 2$). It is this timescale of $\sim 10$ seconds that is probably the best characterization of the collapsar threshold time, the minimum engine working time that will allow the jet to exit the star. As such, when we use the plateau end timescale to constrain progenitor properties below, we use the average properties of the jet (the jet luminosity and opening angle) from this particular high redshift sub-sample.

\subsection{On the Plateau in the Redshift Separated Sub-samples}
A number of studies have shown there is an uptick in the low redshift rate density distribution of GRBs, even when carefully and conservatively accounting for selection effects in the data \citep{Pet15, Yu15, Tsv17, LM17, LR19b, Le20, LR20b, Has24, Nik25}\footnote{Although see \cite{Per16} who note that when a redshift-dependent efficiency of GRB formation from a given progenitor is included in the analysis, this uptick can go away.  Additionally, \cite{LR20b} showed that when accounting for jet opening angle evolution with redshift, the uptick is reduced (although still present).}. Several groups have provided models and arguments for this uptick \citep[e.g.][]{Pet24,LR24}.  In particular they have shown that compact object merger progenitors (e.g. double neutron star mergers and white dwarf-black hole mergers) naturally produce this uptick at low redshifts because of their delay time distributions from formation to merger. 
%\ntp{Who suggested WD-BH mergers producing GRBs? }
%\nlr{i did in the reference above (LR24), from a pop-synth point of view with some back of the envelope energetics and timescales calculated as well, but others before me (also cited above) did simulations and analysis of this particular system as a viable progenitor for long GRBs.}
In other words, the time from the binary formation to the merger event (when the GRB occurs) is on the order of the Hubble time and therefore there is an accumulation of these events at low redshift (relative to their rates at high redshift).   We note the analysis of the uptick in the rate density at low redshifts focused on long-duration GRBs, so this assumes these progenitor systems can produce a long GRB \citep[which appears to be a reasonable assumption, e.g.,][]{Fry99, Zhu22, Gott23, LR24, Chrimes25}. 

Meanwhile, the clear presence of a plateau in the higher redshift sample aligns with the expectation that collapsar progenitors of GRBs favor lower metallicity environments \citep{MW99,YL05,HMM05,Yoon06,WH06}.  In other words, if collapsars are indeed more likely to form in low metallicity environments (as suggested by the studies cited above), then it is natural to expect them to occur more readily at higher redshifts where the average metallicity of the universe is lower. 

\subsection{On the Plateau in the Spectrally Separated Sub-samples}
Similar to what was shown in \cite{Brom13} for the $T_{90}$ distribution, there is no apparent plateau in the $T_{int}$ distribution for our spectrally hard subset of GRBs, shown in Figure~\ref{fig:dNdThardness}, suggesting these GRBs are dominated by non-collapsar progenitors. We might then expect that the hard bursts have on average lower redshifts, aligning with the results mentioned in the previous section that low redshift bursts appear to be dominated by non-collapsar progenitors. 

However, interestingly, the average redshift of both the soft and the hard samples is $\sim 3$. A very big caveat is that our hard sub-sample is relatively small, with only about 25 GRBs.  Within this small sample, there are two GRBs at extremely high redshift (mentioned in the Introduction): GRB120923A at a redshift of $(1+z) = 8.8$ and GRB090429B with $(1+z) = 10.3$.  When those ``outliers'' are removed and we look at only GRBs with $(1+z) < 6$, we find the average redshift of the soft sample is $(1+z) = 2.9$ and the hard sample is $(1+z) = 2.2$. A Student's t-test shows a marginal difference ($\sim 2 \sigma$) between the averages of these two distributions, although again we caution that the numbers in the hard sample are small and so any statistical analysis comparing these distributions is not on firm footing. Nonetheless, it seems clear that there is not a simple correspondence between hard bursts (seemingly dominated by non-collapsar progenitors) and low redshift bursts (also seemingly dominated by non-collapsar progenitors). Indeed this aligns with our analysis in Figure 4 indicating the existence of a class of spectrally soft low-redshift GRBs.

We note that \cite{BP19} and \cite{MN25} have presented compelling evidence for a population of ``fast'' mergers, binary neutron star systems that merge in less than 1 Gyr, which would produce a subset of non-collapsar GRBs at {\em high} redshift.  This could explain the two very high redshift GRBs mentioned above, which are both spectrally hard and short in their intrinsic duration, and the fact that hard GRBs do not show evidence of a plateau in their $dN/dT_{int}$ distribution, regardless of redshift.  We defer a detailed analysis of these GRBs to a future paper. \\

\begin{figure*}[t]
\begin{centering}
\includegraphics[width=0.49\linewidth]{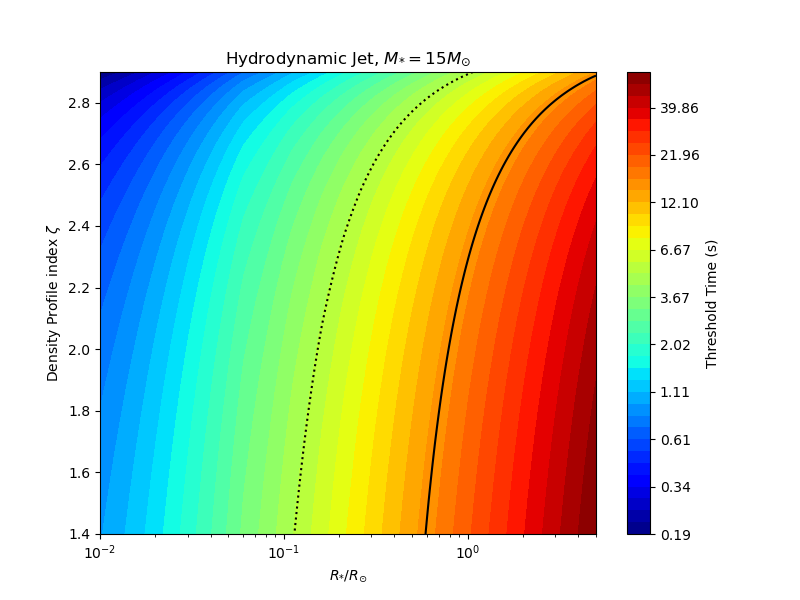}
\includegraphics[width=0.49\linewidth]{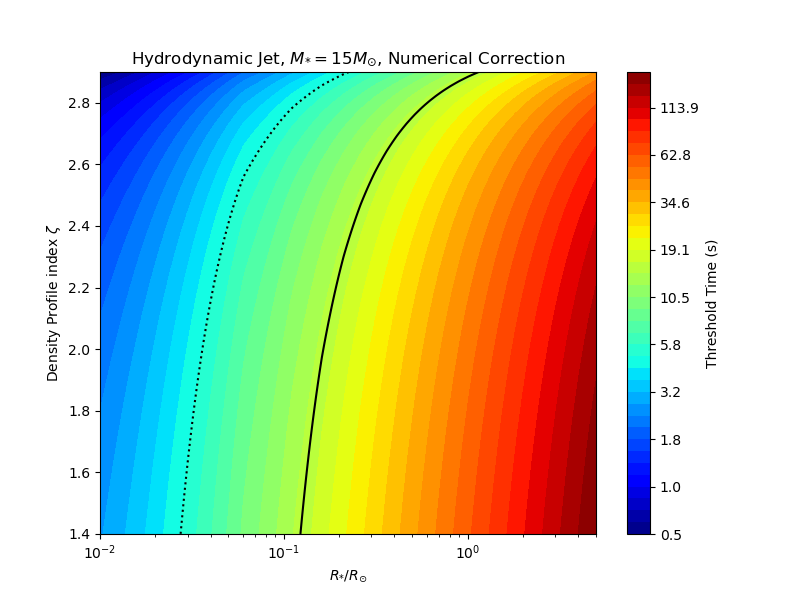}\\
\caption{Threshold time as a function of stellar radius (x-axis) and the stellar density profile index (y-axis) for a star of 15 solar masses. The black dotted and solid lines mark a timescale of 5 seconds and 15 seconds, respectively (corresponding to the range of plateau end times in our $dN/dT_{int}$ distributions). The left panel shows the threshold time scale based on equation~\ref{eq:t_th}, while the right panel applies a correction to this equation based on the numerical simulations of \cite{Harr18}. We used the average (beaming-corrected) jet luminosity and opening angle of our high redshift sample in these calculations.}
\label{fig:Thresholdfull1}
\end{centering}
\end{figure*}

\subsection{Can We Constrain the Properties of Collapsar Progenitors?}

 The end of the plateau in the duration distribution is related to what we defined as the ``threshold time'', $t_{\rm th}$, in \S 1. This is the minimum central engine working time necessary in order for the jet to break out of the star, assuming that the central engine stops abruptly and that the information about the shutoff propagates to the jet head at the speed of light:

\begin{equation}
t_{\rm th}=t_{\rm R}-R/c,
\label{eq:t_th}
\end{equation}

\noindent where $R$ is the stellar radius (including the envelope) and

\begin{equation}
t_{\rm R}=\int_0^R\frac{dr}{\beta_{\rm h}(r)c}
\label{eq:t_R}
\end{equation}

\noindent is the time it takes the jet head to reach the edge of the star.   Another way to think about this timescale is as follows: imagine a jet propagating through the star, and at some point the central engine shuts off abruptly. 
%\rep{The time at which information about the shut off would reach the jet head at the very point the jet reaches the edge of the star}
The shutoff time for which the information about the loss of power catches up with the jet head at the edge of the star is this threshold time.
%equation~\ref{eq:t_th} describes the unique time at which the information about the shut off would reach the jet head at the very point in time that the jet reaches the edge of the star. 
This is the timescale at which we expect the end of the plateau end to appear in the duration distribution of collapsar jets.\\

We now consider this threshold time for different progenitor models, characterized by their density profiles and stellar radii.  
To calculate the threshold time for a given stellar model we need to compute the ratio of the jet energy density to that of the surrounding medium (the rest-mass energy density) at the location of the jet head, defined as $\tilde{L}$ \citep{Matz03,Brom11}:
\begin{equation}
    \tilde{L} = \frac{\rho_{j}(1+4P_{j}/\rho_{j} c^{2})\Gamma_{j}^2}{\rho_{*}},
\end{equation}
\noindent where we assume here a hydrodynamic jet. This ratio sets the velocity of the jet head and, in particular, whether the jet head is relativistic or not. Values of $\tilde{L} \ll (\gg) 1$ correspond to a non-relativistic (relativistic) head velocity. 
To determine this quantity, and therefore measure the jet head speed to be used in equations~\ref{eq:t_th} and ~\ref{eq:t_R}, we do the following:

\begin{itemize}

\item[--]For each stellar profile, we calculate $\tilde{L}(r)$, using equation 3 from \cite{Brom11}:
\begin{equation}
\tilde{L}=\frac{L_j}{\rho_*(r) c^3\Sigma_j(r)},
\end{equation}
where $\rho_*(r)$ is the stellar density at radius $r$ and  $\Sigma_j(r)$ is the jet cross section radius when the jet head is at point $r$. To calculate $\Sigma_j$, we assume that the jet is collimated, $\Sigma_j\simeq\frac{L_j\theta_0^2}{4cp_{\rm c}}$ (their equation 12), where $L_{j}$ is the jet luminosity, $\theta_{o}$ is the half-opening angle of the jet, and $p_{\rm c}$ is the cocoon pressure. 

\item[--]We assume a power-law density profile for our progenitor star over the region the jet propagates, $\rho_*(r) = \rho_{o}(r/r_{o})^{-\zeta}$.  Realistic stellar density profiles do not necessarily follow a single power-law structure but stellar structure simulations of GRB progenitors have shown this can be a good approximation over our region of interest \citep[e.g.][]{WH06,AgDen20, Hal23}. Although the average power-law index can vary substantially across models, it is generally confined between values of $\sim 1.5 - 2.5$.

\item[--]We calculate two functions of $\tilde{L}(r)$ using the non-relativistic, and relativistic expressions for the cocoon pressure $p_{c}$, given in the Appendix of \cite{Brom11}. This results in $\tilde{L}_{NR}\propto(\rho_* r^2)^{-2/3}$(typical for the inner part of the star) and $\tilde{L}_{R}\propto(\rho_* r^2)^{-2/5}$ (typically at the outer parts of the star) for the non-relativistic (NR) and relativistic (R) expressions, respectively. We  locate the radius where $\tilde{L}_{NR}(r_*)=\tilde{L}_{R}(r_*)$. At $r<r_*$ we use the non-relativistic expression, and at $r>r_*$ we used the relativistic expression. 
%This method ensures a smooth transition from a non-relativistic jet propagation
%to a relativistic propagation that occurs at  $\tilde{L}\simeq1$ as expected. 

\item[--] Finally, we obtain the jet head 3-velocity using: 
\begin{equation}
\beta_h(r)\simeq\frac{1}{1+\tilde{L}^{-1/2}}.
\end{equation}
and integrate it using equation~\ref{eq:t_R} to obtain the breakout time of the jet head from the star and substitute the result in equation~\ref{eq:t_th} to obtain the threshold time.\\
\end{itemize}

Our results are shown in Figure~\ref{fig:Thresholdfull1}, which displays contour plots of the threshold time for different values of the density power-law index and stellar radius, assuming a progenitor mass of $15 M_{\odot}$.  We have set $\theta_{o}$ equal to the average value of the jet opening angle in our high redshift sample ($\theta_{o} = \theta_{j} \approx 5.8^{o}$), and set the jet luminosity equal to the average value of the (beaming corrected) jet luminosity in our high redshift sample, $L_{j} \approx 1.9 \times 10^{50} {\rm erg/s}$.\footnote{Note in our soft sample, the average opening angle is $\theta_{j} = 6.3^{o}$, while the average (beaming corrected) jet luminosity is $L_{j} = 0.7 \times 10^{50}$ erg/s.}.
We show similar contour plots over a wider range of parameter space (e.g. different luminosities, progenitor masses, and jet opening angles) in the Appendix. The black dotted and solid lines mark thresholds times of 5 seconds and 15 seconds, respectively.  This corresponds to the range of plateau end times in the $dN/dT$ distributions we find in our collapsar-dominated data samples above.
%\obn{What about the $T_B=2.5 sec$ we found in the complete sample?} 
Again, we argue that the higher redshift sample (orange line in Figure~\ref{fig:dNdT_zsep}) is the subset least contaminated by non-collapsar progenitors, and has an plateau end time of about 10 to 15 seconds. 

The left plot shows the threshold time under the idealized assumptions described above, calculated in equation~\ref{eq:t_th}. We can see for threshold times less than about 15 seconds, GRB collapsar progenitors are constrained to radii less than a solar radius for density profiles shallower than $\zeta \lesssim 2.5$. The right plot shows the same threshold time, but with a numerical correction applied to the equation, as described in \cite{Harr18}. This study found, performing hydrodynamic simulations of a jet propagating through a star with a given density profile, that the simulated threshold time is larger by a factor of $\sim 2$ to $3$ compared to the purely analytic estimate.\footnote{The physical reason behind this is that the jet has a wider effective cross section in the simulations compared to the simple analytic estimate, causing the propagation speed to slow down by a factor of $\sim 2$ to $3$. See \cite{Harr18} for further details.} In this case, progenitor radii are constrained to very small values less than a few tenths of a solar radius.  This is consistent with the size of a carbon oxygen core of a massive star (believed to be the engines of Type Ib/c supernovae), and has interesting implications for the necessity of a binary companion (to strip the envelope to this degree).  We note that if the tip-of-the-iceberg effect, discussed in \S 2, is playing a significant role in the estimates of duration at high redshift (i.e. if we have underestimated these durations), then this radius constraint may be slightly larger.\\

  We note that \cite{Gott22} point out that density profiles that are too steep (with $\zeta > 2$) require over-luminous jets (relative to observed luminosities) to overcome the central high density of the progenitor.  They suggest that only progenitors with $\zeta \leq 2$ will produce GRB jets consistent with observations, under the assumption of a threshold time around 10 seconds and provided the radiative efficiency is relatively high.\footnote{If, instead, one assumes that only a small fraction of jet power is going into the observed radiation, the jet luminosity is much higher than what is inferred from the prompt emission and there is no such a stringent constraint on density index. If this were the case, our jet luminosities would be much higher than the values we are using above and the radius of the progenitor would be constrained to be even smaller.}.\\

Our analysis above is done for a hydrodynamic jet.  This assumption is consistent with the results of \citep{Gott20,Gott21, Gott22} who show significant magnetic energy dissipation in collapsar jets reducing the jet magnetization $\sigma$ to values $\lesssim 0.1$ deep in the stellar core. We point out that magnetic jets are subject to various types of instabilities like the kink instability, which can render them unstable especially in media with a flat density distribution ($\zeta \leq 2$) \citep{Brom16}, resulting in longer breakout times.  Finally, we point out that the value of the threshold time we infer from the intrinsic duration distribution is consistent with the detailed numerical simulations of \cite{Urr25}, of magnetically launched jets propagating through the Wolf-Rayet stellar models of \cite{WH06}.

\section{Summary and Conclusions}

We have re-analyzed the duration distribution of GRBs, following the methods of \cite{Brom12,Brom13}, but using the {\em intrinsic} prompt duration, corrected for cosmological time dilation.  Additionally, we have split our sample into ``low'' and ``high'' redshift, as well as spectrally ``hard'' and ``soft'', sub-samples and found that the distributions differ significantly between these sub-samples. This has important implications for the types of GRBs that dominate at different epochs in the history of our universe.
Our main results are as follows:
\begin{itemize}
    \item We show the presence of a plateau in the {\em intrinsic} duration distribution, $dN/dT_{int}$, where $T_{int}$ is the observed duration corrected for cosmological time dilation, for {\em Swift} GRBs with measured redshifts.  We find the plateau end is shifted to lower durations by a factor of $1/(1+z_{\rm av})$, where $z_{\rm av}$ is the average GRB redshift.  It occurs at around a few seconds, and not 10's of seconds as seen in the {\em observed} prompt duration distribution ($T_{90}$).
    
    \item When we break the sample into lower and higher redshift sub-sets, with a delineating redshift of $(1+z) \sim 2$, we find the strong presence of a plateau in the high redshift sample, with an end at around $10$ to $15$ seconds and no evidence of a plateau in the low redshift sample of GRBs. 
    We suggest this aligns with recent studies that have claimed that the low redshift sample is dominated by non-collapsar (potentially compact object merger) progenitors.
    %\obn{here the inconsistency between the plateau end in the intrinsic distribution of the complete sample and the high redshift sample is evident again.}.

    \item When our sample is split into spectrally harder and softer sub-samples, the spectrally hard sample exhibits no plateau in the distribution, while the spectrally soft sample exhibits a clear plateau.  This confirms what was found in \cite{Brom13} for the $T_{90}$ distribution. Furthermore, although spectrally hard GRBs appear to be dominated by non-collapsars across redshifts, spectrally soft GRBs show a dichotomy between low and high redshift sub-samples. When the spectrally soft sample is separated into low and high redshift groups, the plateau only appears in the high redshift sample while there is no plateau in the spectrally soft low-redshift sample, shown in Figure~\ref{fig:dNdThardnessredsep}. This suggests {\em there exists a significant population of spectrally soft non-collapsar progenitors at low redshift}. 

    \item  Connecting the end time of the plateau to the ``threshold time'' (the minimum central engine required to push the jet out of the star) of a typical hydrodynamic collapsar jet, we use our results to constrain the radius of the progenitor star. We show that, when realistic assumptions about the jet head cross section and propagation velocity are accounted for, it must be less than only a few {\em tenths} of a solar radius, consistent with the size of a carbon-oxygen core of a massive star.  Although it has long been argued that GRBs should come from stripped envelope stars, these envelopes can in some cases be optically thick and this analysis puts relatively strong limits on the extent and density of the wind envelope around GRB progenitors.
    %\obn{Can we put numbers to this claim?}\nlr{Worth discussing.}
 
\end{itemize}

 As we continue to discover interesting examples of GRBs that upend our traditional picture of classifying their progenitors according to whether the observed prompt gamma-ray burst is long/soft or short/hard, looking for other signatures in the data that may help clarify the progenitor paradigm is more important than ever.  Telescopes like the {\em Einstein Probe} have, for example, discovered X-ray transients at rates consistent with GRB event rates and redshift distributions \citep{Oconn25, Gao25}. {\em SVOM} has begun detecting short duration GRBs out to high redshifts \citep{Dimple25}. We soon  may be able to obtain better statistics on supernova counterparts or currently small sub-populations like low luminosity GRBs \citep{Vir09}, which will help untangle the GRB progenitor paradigm. Meanwhile, an emerging multi-messenger astrophysics community is coordinating efforts to optimize the follow-up of GRBs across detector capabilities. With the advent of these new observatories and observing strategies, we are poised to access the keys to unlocking the mystery behind these most extreme events in our universe.
% \obn{the last sentence is way too long, we need to break it into 3 sentences at least. I would also consider splitting the first sentence to two, but up to you.}. 

\begin{acknowledgments}
We thank the anonymous referee for their review, which led to improvements in our manuscript.  N.L.-R. thanks Chris Matzner, Fabio De Colle, Chris Thompson, and James Leung for interesting conversations related to this work. This work was supported by the U.~S. Department of Energy through Los Alamos National Laboratory (LANL).  LANL is operated by Triad National Security, LLC, for the National Nuclear Security Administration of U.S. Department of Energy (Contract No. 89233218CNA000001), LA-UR-25-31112.   
%N.L.-R. acknowledges support from Laboratory Directed Research and Development program, LANL project numbers 20230115ER and 20230217ER. 
O.B. was supported by an ISF grant 2067/22, a BSF grant 2024297 and NSF-BSF grants 2020747 \& 2024788. T.P. acknowledges support from an Advanced ERC grant MultiJets, ISF grant 2126 and the Simon foundation SCEECS collaboration.  \\
\end{acknowledgments}

\begin{contribution}

All authors contributed equally to the collaboration.

\end{contribution}

%\bibliography{refs}
%\bibliographystyle{aasjournalv7}
\bibliographystyle{aasjournal}
\bibliography{refs}

@ARTICLE{Kasliwal2017,
       author = {{Kasliwal}, M.~M. and {Nakar}, E. and {Singer}, L.~P. and {Kaplan}, D.~L. and {Cook}, D.~O. and {Van Sistine}, A. and {Lau}, R.~M. and {Fremling}, C. and {Gottlieb}, O. and {Jencson}, J.~E. and {Adams}, S.~M. and {Feindt}, U. and {Hotokezaka}, K. and {Ghosh}, S. and {Perley}, D.~A. and {Yu}, P.-C. and {Piran}, T. and {Allison}, J.~R. and {Anupama}, G.~C. and {Balasubramanian}, A. and {Bannister}, K.~W. and {Bally}, J. and {Barnes}, J. and {Barway}, S. and {Bellm}, E. and {Bhalerao}, V. and {Bhattacharya}, D. and {Blagorodnova}, N. and {Bloom}, J.~S. and {Brady}, P.~R. and {Cannella}, C. and {Chatterjee}, D. and {Cenko}, S.~B. and {Cobb}, B.~E. and {Copperwheat}, C. and {Corsi}, A. and {De}, K. and {Dobie}, D. and {Emery}, S.~W.~K. and {Evans}, P.~A. and {Fox}, O.~D. and {Frail}, D.~A. and {Frohmaier}, C. and {Goobar}, A. and {Hallinan}, G. and {Harrison}, F. and {Helou}, G. and {Hinderer}, T. and {Ho}, A.~Y.~Q. and {Horesh}, A. and {Ip}, W.-H. and {Itoh}, R. and {Kasen}, D. and {Kim}, H. and {Kuin}, N.~P.~M. and {Kupfer}, T. and {Lynch}, C. and {Madsen}, K. and {Mazzali}, P.~A. and {Miller}, A.~A. and {Mooley}, K. and {Murphy}, T. and {Ngeow}, C.-C. and {Nichols}, D. and {Nissanke}, S. and {Nugent}, P. and {Ofek}, E.~O. and {Qi}, H. and {Quimby}, R.~M. and {Rosswog}, S. and {Rusu}, F. and {Sadler}, E.~M. and {Schmidt}, P. and {Sollerman}, J. and {Steele}, I. and {Williamson}, A.~R. and {Xu}, Y. and {Yan}, L. and {Yatsu}, Y. and {Zhang}, C. and {Zhao}, W.},
        title = "{Illuminating gravitational waves: A concordant picture of photons from a neutron star merger}",
      journal = {Science},
         year = 2017,
        month = dec,
       volume = {358},
       number = {6370},
        pages = {1559-1565},
          doi = {10.1126/science.aap9455},
archivePrefix = {arXiv},
       eprint = {1710.05436},
 primaryClass = {astro-ph.HE},
       adsurl = {https://ui.adsabs.harvard.edu/abs/2017Sci...358.1559K}
}

@ARTICLE{NP02b,
       author = {{Nakar}, Ehud and {Piran}, Tsvi},
        title = "{Temporal properties of short gamma-ray bursts}",
      journal = {\mnras},
         year = 2002,
        month = mar,
       volume = {330},
       number = {4},
        pages = {920-926},
          doi = {10.1046/j.1365-8711.2002.05136.x},
archivePrefix = {arXiv},
       eprint = {astro-ph/0103192},
 primaryClass = {astro-ph},
       adsurl = {https://ui.adsabs.harvard.edu/abs/2002MNRAS.330..920N}
}

@ARTICLE{Littlejohns13,
       author = {{Littlejohns}, O.~M. and {Tanvir}, N.~R. and {Willingale}, R. and {Evans}, P.~A. and {O'Brien}, P.~T. and {Levan}, A.~J.},
        title = "{Are gamma-ray bursts the same at high redshift and low redshift?}",
      journal = {\mnras},
         year = 2013,
        month = dec,
       volume = {436},
       number = {4},
        pages = {3640-3655},
          doi = {10.1093/mnras/stt1841},
archivePrefix = {arXiv},
       eprint = {1309.7045},
 primaryClass = {astro-ph.HE},
       adsurl = {https://ui.adsabs.harvard.edu/abs/2013MNRAS.436.3640L}
}

@ARTICLE{Moss26,
       author = {{Moss}, Michael J. and {Lien}, Amy Y. and {Cenko}, S. Bradley and {Guiriec}, Sylvain and {Markwardt}, Craig B.},
        title = "{How Distance Affects GRB Prompt Emission Measurements}",
      journal = {arXiv e-prints},
         year = 2026,
        month = feb,
          eid = {arXiv:2602.03032},
        pages = {arXiv:2602.03032},
          doi = {10.48550/arXiv.2602.03032},
archivePrefix = {arXiv},
       eprint = {2602.03032},
 primaryClass = {astro-ph.HE},
       adsurl = {https://ui.adsabs.harvard.edu/abs/2026arXiv260203032M}
}

@ARTICLE{Wang20,
       author = {{Wang}, Feifei and {Zou}, Yuan-Chuan and {Liu}, Fuxiang and {Liao}, Bin and {Liu}, Yu and {Chai}, Yating and {Xia}, Lei},
        title = "{A Comprehensive Statistical Study of Gamma-Ray Bursts}",
      journal = {\apj},
         year = 2020,
        month = apr,
       volume = {893},
       number = {1},
          eid = {77},
        pages = {77},
          doi = {10.3847/1538-4357/ab0a86},
archivePrefix = {arXiv},
       eprint = {1902.05489},
 primaryClass = {astro-ph.HE},
       adsurl = {https://ui.adsabs.harvard.edu/abs/2020ApJ...893...77W}
}

@ARTICLE{Moss22,
       author = {{Moss}, Michael and {Lien}, Amy and {Guiriec}, Sylvain and {Cenko}, S. Bradley and {Sakamoto}, Takanori},
        title = "{Instrumental Tip-of-the-iceberg Effects on the Prompt Emission of Swift/BAT Gamma-ray Bursts}",
      journal = {\apj},
         year = 2022,
        month = mar,
       volume = {927},
       number = {2},
          eid = {157},
        pages = {157},
          doi = {10.3847/1538-4357/ac4d94},
archivePrefix = {arXiv},
       eprint = {2111.13392},
 primaryClass = {astro-ph.HE},
       adsurl = {https://ui.adsabs.harvard.edu/abs/2022ApJ...927..157M}
}

@ARTICLE{Oconn25,
       author = {{O'Connor}, Brendan and {Beniamini}, Paz and {Troja}, Eleonora and {Busmann}, Malte and {Dichiara}, Simone and {Gill}, Ramandeep and {Granot}, Jonathan and {Moss}, Michael and {Hall}, Xander and {Palmese}, Antonella and {Passaleva}, Niccolo and {Yang}, Yu-Han},
        title = "{The redshift distribution of Einstein Probe transients supports their relation to gamma-ray bursts}",
      journal = {arXiv e-prints},
         year = 2025,
        month = sep,
          eid = {arXiv:2509.07141},
        pages = {arXiv:2509.07141},
          doi = {10.48550/arXiv.2509.07141},
archivePrefix = {arXiv},
       eprint = {2509.07141},
 primaryClass = {astro-ph.HE},
       adsurl = {https://ui.adsabs.harvard.edu/abs/2025arXiv250907141O}
}

@ARTICLE{Gao25,
       author = {{Gao}, Hao-Xuan and {Geng}, Jin-Jun and {Liang}, Yi-Fang and {Sun}, Hui and {Xu}, Fan and {Wu}, Xue-Feng and {Huang}, Yong-Feng and {Dai}, Zi-Gao and {Yuan}, Wei-Min},
        title = "{The Soft X-Ray Aspect of Gamma-Ray Bursts in the Einstein Probe Era}",
      journal = {\apj},
         year = 2025,
        month = jun,
       volume = {986},
       number = {1},
          eid = {106},
        pages = {106},
          doi = {10.3847/1538-4357/adceb1},
archivePrefix = {arXiv},
       eprint = {2410.21687},
 primaryClass = {astro-ph.HE},
       adsurl = {https://ui.adsabs.harvard.edu/abs/2025ApJ...986..106G}
}

@ARTICLE{Dimple25,
       author = {{Dimple} and {Gompertz}, B.~P. and {Levan}, A.~J. and {Malesani}, D.~B. and {Laskar}, T. and {Bala}, S. and {Chrimes}, A.~A. and {Heintz}, K. and {Izzo}, L. and {Lamb}, G.~P. and {O'Neill}, D. and {Palmerio}, J.~T. and {Saccardi}, A. and {Anderson}, G.~E. and {De Barra}, C. and {Huang}, Y. and {Kumar}, A. and {Li}, H. and {McBreen}, S. and {Mukherjee}, O. and {Oates}, S.~R. and {Pathak}, U. and {Qiu}, Y. and {Roberts}, O.~J. and {Sonawane}, R. and {Veres}, P. and {Ackley}, K. and {Han}, X. and {Julakanti}, Y. and {Wang}, J. and {D'Avanzo}, P. and {Martin-Carrillo}, A. and {Ravasio}, M.~E. and {Rossi}, A. and {Tanvir}, N.~R. and {Anderson}, J.~P. and {Arabsalmani}, M. and {Belkin}, S. and {Breton}, R.~P. and {Brivio}, R. and {Burns}, E. and {Casares}, J. and {Campana}, S. and {Chastain}, S.~I. and {D'Elia}, V. and {Dhillon}, V.~S. and {Dyer}, M.~J. and {Fynbo}, J.~P.~U. and {Galloway}, D.~K. and {Gulati}, A. and {Godson}, B. and {Goodwin}, A.~J. and {Gromadzki}, M. and {Hartmann}, D.~H. and {Jakobsson}, P. and {Killestein}, T.~L. and {Kotak}, R. and {Leung}, J.~K. and {Lyman}, J.~D. and {Melandri}, A. and {Mattila}, S. and {McGee}, S. and {Morley}, C. and {Mukherjee}, T. and {M{\"u}ller-Bravo}, T.~E. and {Noysena}, K. and {Nuttall}, L.~K. and {O'Brien}, P. and {De Pasquale}, M. and {Pignata}, G. and {Pollacco}, D. and {Pugliese}, G. and {Ramsay}, G. and {Sahu}, A. and {Salvaterra}, R. and {Schady}, P. and {Schneider}, B. and {Steeghs}, D. and {Starling}, R.~L.~C. and {Tsalapatas}, K. and {Ulaczyk}, K. and {van der Horst}, A.~J. and {Wang}, C. and {Wiersema}, K. and {Worssam}, I. and {Wortley}, M.~E. and {Xiong}, S. and {Zafar}, T.},
        title = "{GRB 241105A: a test case for GRB classification and rapid r-process nucleosynthesis channels}",
      journal = {\mnras},
         year = 2025,
        month = nov,
       volume = {544},
       number = {1},
        pages = {548-571},
          doi = {10.1093/mnras/staf1574},
archivePrefix = {arXiv},
       eprint = {2507.15940},
 primaryClass = {astro-ph.HE},
       adsurl = {https://ui.adsabs.harvard.edu/abs/2025MNRAS.544..548D}
}

@ARTICLE{Urr25,
       author = {{Urrutia}, Gerardo and {Janiuk}, Agnieszka and {Olivares}, Hector},
        title = "{Numerical simulations of jet launching and breakout from collapsars}",
      journal = {arXiv e-prints},
         year = 2025,
        month = jul,
          eid = {arXiv:2507.10231},
        pages = {arXiv:2507.10231},
          doi = {10.48550/arXiv.2507.10231},
archivePrefix = {arXiv},
       eprint = {2507.10231},
 primaryClass = {astro-ph.HE},
       adsurl = {https://ui.adsabs.harvard.edu/abs/2025arXiv250710231U}
}

@ARTICLE{Ris25,
       author = {{Risti{\'c}}, Marko and {Barker}, Brandon L. and {Cupp}, Samuel and {Gross}, Axel and {Lloyd-Ronning}, Nicole and {Korobkin}, Oleg and {Miller}, Jonah M. and {Mumpower}, Matthew R.},
        title = "{Kilonovae and Long-duration Gamma-ray Bursts}",
      journal = {arXiv e-prints},
         year = 2025,
        month = sep,
          eid = {arXiv:2509.03003},
        pages = {arXiv:2509.03003},
          doi = {10.48550/arXiv.2509.03003},
archivePrefix = {arXiv},
       eprint = {2509.03003},
 primaryClass = {astro-ph.HE},
       adsurl = {https://ui.adsabs.harvard.edu/abs/2025arXiv250903003R}
}

@ARTICLE{MN25,
       author = {{Maoz}, Dan and {Nakar}, Ehud},
        title = "{The Neutron Star Merger Delay-time Distribution, R-process ``Knees,'' and the Metal Budget of the Galaxy}",
      journal = {\apj},
         year = 2025,
        month = apr,
       volume = {982},
       number = {2},
          eid = {179},
        pages = {179},
          doi = {10.3847/1538-4357/ada3bd},
archivePrefix = {arXiv},
       eprint = {2406.08630},
 primaryClass = {astro-ph.HE},
       adsurl = {https://ui.adsabs.harvard.edu/abs/2025ApJ...982..179M}
}

@ARTICLE{BP19,
       author = {{Beniamini}, Paz and {Piran}, Tsvi},
        title = "{The Gravitational waves merger time distribution of binary neutron star systems}",
      journal = {\mnras},
         year = 2019,
        month = aug,
       volume = {487},
       number = {4},
        pages = {4847-4854},
          doi = {10.1093/mnras/stz1589},
archivePrefix = {arXiv},
       eprint = {1903.11614},
 primaryClass = {astro-ph.HE},
       adsurl = {https://ui.adsabs.harvard.edu/abs/2019MNRAS.487.4847B}
}

@ARTICLE{MP17,
       author = {{Moharana}, Reetanjali and {Piran}, Tsvi},
        title = "{Observational evidence for mass ejection accompanying short gamma-ray bursts}",
      journal = {\mnras},
         year = 2017,
        month = nov,
       volume = {472},
       number = {1},
        pages = {L55-L59},
          doi = {10.1093/mnrasl/slx131},
archivePrefix = {arXiv},
       eprint = {1705.02598},
 primaryClass = {astro-ph.HE},
       adsurl = {https://ui.adsabs.harvard.edu/abs/2017MNRAS.472L..55M}
}

@ARTICLE{Cuc11,
       author = {{Cucchiara}, A. and {Levan}, A.~J. and {Fox}, D.~B. and {Tanvir}, N.~R. and {Ukwatta}, T.~N. and {Berger}, E. and {Kr{\"u}hler}, T. and {K{\"u}pc{\"u} Yolda{\c{s}}}, A. and {Wu}, X.~F. and {Toma}, K. and {Greiner}, J. and {Olivares}, F.~E. and {Rowlinson}, A. and {Amati}, L. and {Sakamoto}, T. and {Roth}, K. and {Stephens}, A. and {Fritz}, Alexander and {Fynbo}, J.~P.~U. and {Hjorth}, J. and {Malesani}, D. and {Jakobsson}, P. and {Wiersema}, K. and {O'Brien}, P.~T. and {Soderberg}, A.~M. and {Foley}, R.~J. and {Fruchter}, A.~S. and {Rhoads}, J. and {Rutledge}, R.~E. and {Schmidt}, B.~P. and {Dopita}, M.~A. and {Podsiadlowski}, P. and {Willingale}, R. and {Wolf}, C. and {Kulkarni}, S.~R. and {D'Avanzo}, P.},
        title = "{A Photometric Redshift of z \raisebox{-0.5ex}\textasciitilde 9.4 for GRB 090429B}",
      journal = {\apj},
         year = 2011,
        month = jul,
       volume = {736},
       number = {1},
          eid = {7},
        pages = {7},
          doi = {10.1088/0004-637X/736/1/7},
archivePrefix = {arXiv},
       eprint = {1105.4915},
 primaryClass = {astro-ph.CO},
       adsurl = {https://ui.adsabs.harvard.edu/abs/2011ApJ...736....7C}
}

@ARTICLE{Tan18,
       author = {{Tanvir}, N.~R. and {Laskar}, T. and {Levan}, A.~J. and {Perley}, D.~A. and {Zabl}, J. and {Fynbo}, J.~P.~U. and {Rhoads}, J. and {Cenko}, S.~B. and {Greiner}, J. and {Wiersema}, K. and {Hjorth}, J. and {Cucchiara}, A. and {Berger}, E. and {Bremer}, M.~N. and {Cano}, Z. and {Cobb}, B.~E. and {Covino}, S. and {D'Elia}, V. and {Fong}, W. and {Fruchter}, A.~S. and {Goldoni}, P. and {Hammer}, F. and {Heintz}, K.~E. and {Jakobsson}, P. and {Kann}, D.~A. and {Kaper}, L. and {Klose}, S. and {Knust}, F. and {Kr{\"u}hler}, T. and {Malesani}, D. and {Misra}, K. and {Nicuesa Guelbenzu}, A. and {Pugliese}, G. and {S{\'a}nchez-Ram{\'\i}rez}, R. and {Schulze}, S. and {Stanway}, E.~R. and {de Ugarte Postigo}, A. and {Watson}, D. and {Wijers}, R.~A.~M.~J. and {Xu}, D.},
        title = "{The Properties of GRB 120923A at a Spectroscopic Redshift of z {\ensuremath{\approx}} 7.8}",
      journal = {\apj},
         year = 2018,
        month = oct,
       volume = {865},
       number = {2},
          eid = {107},
        pages = {107},
          doi = {10.3847/1538-4357/aadba9},
archivePrefix = {arXiv},
       eprint = {1703.09052},
 primaryClass = {astro-ph.HE},
       adsurl = {https://ui.adsabs.harvard.edu/abs/2018ApJ...865..107T}
}

@ARTICLE{NPP92,
       author = {{Narayan}, Ramesh and {Paczynski}, Bohdan and {Piran}, Tsvi},
        title = "{Gamma-Ray Bursts as the Death Throes of Massive Binary Stars}",
      journal = {\apjl},
         year = 1992,
        month = aug,
       volume = {395},
        pages = {L83},
          doi = {10.1086/186493},
archivePrefix = {arXiv},
       eprint = {astro-ph/9204001},
 primaryClass = {astro-ph},
       adsurl = {https://ui.adsabs.harvard.edu/abs/1992ApJ...395L..83N}
}

@ARTICLE{Fong2010,
   author = {{Fong}, W. and {Berger}, E. and {Fox}, D.~B.},
    title = "{Hubble Space Telescope Observations of Short Gamma-Ray Burst Host Galaxies: Morphologies, Offsets, and Local Environments}",
  journal = {\apj},
archivePrefix = "arXiv",
   eprint = {0909.1804},
 primaryClass = "astro-ph.HE",
     year = 2010,
    month = jan,
   volume = 708,
    pages = {9-25},
      doi = {10.1088/0004-637X/708/1/9},
   adsurl = {http://adsabs.harvard.edu/abs/2010ApJ...708....9F}
}

@ARTICLE{Sod06,
       author = {{Soderberg}, A.~M. and {Kulkarni}, S.~R. and {Nakar}, E. and {Berger}, E. and {Cameron}, P.~B. and {Fox}, D.~B. and {Frail}, D. and {Gal-Yam}, A. and {Sari}, R. and {Cenko}, S.~B. and {Kasliwal}, M. and {Chevalier}, R.~A. and {Piran}, T. and {Price}, P.~A. and {Schmidt}, B.~P. and {Pooley}, G. and {Moon}, D. -S. and {Penprase}, B.~E. and {Ofek}, E. and {Rau}, A. and {Gehrels}, N. and {Nousek}, J.~A. and {Burrows}, D.~N. and {Persson}, S.~E. and {McCarthy}, P.~J.},
        title = "{Relativistic ejecta from X-ray flash XRF 060218 and the rate of cosmic explosions}",
      journal = {\nat},
         year = 2006,
        month = aug,
       volume = {442},
       number = {7106},
        pages = {1014-1017},
          doi = {10.1038/nature05087},
archivePrefix = {arXiv},
       eprint = {astro-ph/0604389},
 primaryClass = {astro-ph},
       adsurl = {https://ui.adsabs.harvard.edu/abs/2006Natur.442.1014S}
}

@ARTICLE{Luo23,
       author = {{Luo}, Jia-Wei and {Wang}, Fei-Fei and {Zhu-Ge}, Jia-Ming and {Li}, Ye and {Zou}, Yuan-Chuan and {Zhang}, Bing},
        title = "{Identifying the Physical Origin of Gamma-Ray Bursts with Supervised Machine Learning}",
      journal = {\apj},
         year = 2023,
        month = dec,
       volume = {959},
       number = {1},
          eid = {44},
        pages = {44},
          doi = {10.3847/1538-4357/ad03ec},
archivePrefix = {arXiv},
       eprint = {2211.16451},
 primaryClass = {astro-ph.HE},
       adsurl = {https://ui.adsabs.harvard.edu/abs/2023ApJ...959...44L}
}

@ARTICLE{Chen24b,
       author = {{Chen}, Jia-Ming and {Zhu}, Ke-Rui and {Peng}, Zhao-Yang and {Zhang}, Li},
        title = "{Unsupervised machine learning classification of Fermi gamma-ray bursts using spectral parameters}",
      journal = {\mnras},
         year = 2024,
        month = jan,
       volume = {527},
       number = {2},
        pages = {4272-4284},
          doi = {10.1093/mnras/stad3407},
       adsurl = {https://ui.adsabs.harvard.edu/abs/2024MNRAS.527.4272C}
}

@ARTICLE{Esp25,
       author = {{Espinoza}, Sharleen N. and {Lloyd-Ronning}, Nicole M. and {Negro}, Michela and {Cheng}, Roseanne M. and {Cibrario}, Nicol{\'o}},
        title = "{Mapping Gamma-Ray Bursts: Distinguishing Progenitor Systems Through Machine Learning}",
      journal = {Research Notes of the American Astronomical Society},
         year = 2025,
        month = sep,
       volume = {9},
       number = {9},
          eid = {239},
        pages = {239},
          doi = {10.3847/2515-5172/ae031c},
archivePrefix = {arXiv},
       eprint = {2508.20214},
 primaryClass = {astro-ph.HE},
       adsurl = {https://ui.adsabs.harvard.edu/abs/2025RNAAS...9..239E}
}

@ARTICLE{Garc24,
       author = {{Garcia-Cifuentes}, Keneth and {Becerra}, Rosa and {De Colle}, Fabio},
        title = "{ClassiPyGRB: Machine Learning-Based Classification and Visualization of Gamma Ray Bursts using t-SNE}",
      journal = {The Journal of Open Source Software},
         year = 2024,
        month = apr,
       volume = {9},
       number = {96},
          eid = {5923},
        pages = {5923},
          doi = {10.21105/joss.05923},
archivePrefix = {arXiv},
       eprint = {2404.06439},
 primaryClass = {astro-ph.HE},
       adsurl = {https://ui.adsabs.harvard.edu/abs/2024JOSS....9.5923G}
}

@ARTICLE{Metz08,
       author = {{Metzger}, B.~D. and {Quataert}, E. and {Thompson}, T.~A.},
        title = "{Short-duration gamma-ray bursts with extended emission from protomagnetar spin-down}",
      journal = {\mnras},
         year = 2008,
        month = apr,
       volume = {385},
       number = {3},
        pages = {1455-1460},
          doi = {10.1111/j.1365-2966.2008.12923.x},
archivePrefix = {arXiv},
       eprint = {0712.1233},
 primaryClass = {astro-ph},
       adsurl = {https://ui.adsabs.harvard.edu/abs/2008MNRAS.385.1455M}
}

@ARTICLE{Dimple24,
       author = {{Dimple} and {Misra}, Kuntal and {Yadav}, Lallan},
        title = "{Investigating High Redshift Short GRBs: Signatures of Collapsars?}",
      journal = {Bulletin de la Societe Royale des Sciences de Liege},
         year = 2024,
        month = jun,
       volume = {93},
       number = {2},
        pages = {670-682},
          doi = {10.25518/0037-9565.11833},
archivePrefix = {arXiv},
       eprint = {2309.12788},
 primaryClass = {astro-ph.HE},
       adsurl = {https://ui.adsabs.harvard.edu/abs/2024BSRSL..93..670D}
}

@ARTICLE{Dimple22,
       author = {{Dimple} and {Misra}, K. and {Kann}, D.~A. and {Arun}, K.~G. and {Ghosh}, A. and {Gupta}, R. and {Resmi}, L. and {Ag{\"u}{\'\i} Fern{\'a}ndez}, J.~F. and {Th{\"o}ne}, C.~C. and {de Ugarte Postigo}, A. and {Pandey}, S.~B. and {Yadav}, L.},
        title = "{Multiwavelength analysis of short GRB 201221D and its comparison with other high \& low redshift short GRBs}",
      journal = {\mnras},
         year = 2022,
        month = oct,
       volume = {516},
       number = {1},
        pages = {1-12},
          doi = {10.1093/mnras/stac2162},
archivePrefix = {arXiv},
       eprint = {2206.08947},
 primaryClass = {astro-ph.HE},
       adsurl = {https://ui.adsabs.harvard.edu/abs/2022MNRAS.516....1D}
}

@ARTICLE{Kan15,
       author = {{Kaneko}, Y. and {Bostanc{\i}}, Z.~F. and {G{\"o}{\u{g}}{\"u}{\c{s}}}, E. and {Lin}, L.},
        title = "{Short gamma-ray bursts with extended emission observed with Swift/BAT and Fermi/GBM}",
      journal = {\mnras},
         year = 2015,
        month = sep,
       volume = {452},
       number = {1},
        pages = {824-837},
          doi = {10.1093/mnras/stv1286},
archivePrefix = {arXiv},
       eprint = {1506.05899},
 primaryClass = {astro-ph.HE},
       adsurl = {https://ui.adsabs.harvard.edu/abs/2015MNRAS.452..824K}
}

@ARTICLE{Guet07,
       author = {{Guetta}, Dafne and {Della Valle}, Massimo},
        title = "{On the Rates of Gamma-Ray Bursts and Type Ib/c Supernovae}",
      journal = {\apjl},
         year = 2007,
        month = mar,
       volume = {657},
       number = {2},
        pages = {L73-L76},
          doi = {10.1086/511417},
archivePrefix = {arXiv},
       eprint = {astro-ph/0612194},
 primaryClass = {astro-ph},
       adsurl = {https://ui.adsabs.harvard.edu/abs/2007ApJ...657L..73G}
}

@ARTICLE{Ross22,
       author = {{Rossi}, A. and {Rothberg}, B. and {Palazzi}, E. and {Kann}, D.~A. and {D'Avanzo}, P. and {Amati}, L. and {Klose}, S. and {Perego}, A. and {Pian}, E. and {Guidorzi}, C. and {Pozanenko}, A.~S. and {Savaglio}, S. and {Stratta}, G. and {Agapito}, G. and {Covino}, S. and {Cusano}, F. and {D'Elia}, V. and {De Pasquale}, M. and {Della Valle}, M. and {Kuhn}, O. and {Izzo}, L. and {Loffredo}, E. and {Masetti}, N. and {Melandri}, A. and {Minaev}, P.~Y. and {Guelbenzu}, A. Nicuesa and {Paris}, D. and {Paiano}, S. and {Plantet}, C. and {Rossi}, F. and {Salvaterra}, R. and {Schulze}, S. and {Veillet}, C. and {Volnova}, A.~A.},
        title = "{The Peculiar Short-duration GRB 200826A and Its Supernova}",
      journal = {\apj},
         year = 2022,
        month = jun,
       volume = {932},
       number = {1},
          eid = {1},
        pages = {1},
          doi = {10.3847/1538-4357/ac60a2},
archivePrefix = {arXiv},
       eprint = {2105.03829},
 primaryClass = {astro-ph.HE},
       adsurl = {https://ui.adsabs.harvard.edu/abs/2022ApJ...932....1R}
}

@ARTICLE{NP02,
       author = {{Nakar}, Ehud and {Piran}, Tsvi},
        title = "{Time-scales in long gamma-ray bursts}",
      journal = {\mnras},
         year = 2002,
        month = mar,
       volume = {331},
       number = {1},
        pages = {40-44},
          doi = {10.1046/j.1365-8711.2002.05158.x},
archivePrefix = {arXiv},
       eprint = {astro-ph/0103210},
 primaryClass = {astro-ph},
       adsurl = {https://ui.adsabs.harvard.edu/abs/2002MNRAS.331...40N}
}

@ARTICLE{Pir04,
       author = {{Piran}, Tsvi},
        title = "{The physics of gamma-ray bursts}",
      journal = {Reviews of Modern Physics},
         year = 2004,
        month = oct,
       volume = {76},
       number = {4},
        pages = {1143-1210},
          doi = {10.1103/RevModPhys.76.1143},
archivePrefix = {arXiv},
       eprint = {astro-ph/0405503},
 primaryClass = {astro-ph},
       adsurl = {https://ui.adsabs.harvard.edu/abs/2004RvMP...76.1143P}
}

@ARTICLE{Fong2013,
   author = {{Fong}, W. and {Berger}, E.},
    title = "{The Locations of Short Gamma-Ray Bursts as Evidence for Compact Object Binary Progenitors}",
  journal = {\apj},
archivePrefix = "arXiv",
   eprint = {1307.0819},
 primaryClass = "astro-ph.HE",
     year = 2013,
    month = oct,
   volume = 776,
      eid = {18},
    pages = {18},
      doi = {10.1088/0004-637X/776/1/18},
   adsurl = {http://adsabs.harvard.edu/abs/2013ApJ...776...18F}
}

@ARTICLE{Brom12,
       author = {{Bromberg}, Omer and {Nakar}, Ehud and {Piran}, Tsvi and {Sari}, Re'em},
        title = "{An Observational Imprint of the Collapsar Model of Long Gamma-Ray Bursts}",
      journal = {\apj},
         year = 2012,
        month = apr,
       volume = {749},
       number = {2},
          eid = {110},
        pages = {110},
          doi = {10.1088/0004-637X/749/2/110},
archivePrefix = {arXiv},
       eprint = {1111.2990},
 primaryClass = {astro-ph.HE},
       adsurl = {https://ui.adsabs.harvard.edu/abs/2012ApJ...749..110B}
}

@ARTICLE{Gott20,
       author = {{Gottlieb}, Ore and {Bromberg}, Omer and {Singh}, Chandra B. and {Nakar}, Ehud},
        title = "{The structure of weakly magnetized {\ensuremath{\gamma}}-ray burst jets}",
      journal = {\mnras},
         year = 2020,
        month = nov,
       volume = {498},
       number = {3},
        pages = {3320-3333},
          doi = {10.1093/mnras/staa2567},
archivePrefix = {arXiv},
       eprint = {2007.11590},
 primaryClass = {astro-ph.HE},
       adsurl = {https://ui.adsabs.harvard.edu/abs/2020MNRAS.498.3320G}
}

@ARTICLE{Gott21,
       author = {{Gottlieb}, Ore and {Bromberg}, Omer and {Levinson}, Amir and {Nakar}, Ehud},
        title = "{Intermittent mildly magnetized jets as the source of GRBs}",
      journal = {\mnras},
         year = 2021,
        month = jul,
       volume = {504},
       number = {3},
        pages = {3947-3955},
          doi = {10.1093/mnras/stab1068},
archivePrefix = {arXiv},
       eprint = {2102.00005},
 primaryClass = {astro-ph.HE},
       adsurl = {https://ui.adsabs.harvard.edu/abs/2021MNRAS.504.3947G}
}

@ARTICLE{Gott22,
       author = {{Gottlieb}, Ore and {Lalakos}, Aretaios and {Bromberg}, Omer and {Liska}, Matthew and {Tchekhovskoy}, Alexander},
        title = "{Black hole to breakout: 3D GRMHD simulations of collapsar jets reveal a wide range of transients}",
      journal = {\mnras},
         year = 2022,
        month = mar,
       volume = {510},
       number = {4},
        pages = {4962-4975},
          doi = {10.1093/mnras/stab3784},
archivePrefix = {arXiv},
       eprint = {2109.14619},
 primaryClass = {astro-ph.HE},
       adsurl = {https://ui.adsabs.harvard.edu/abs/2022MNRAS.510.4962G}
}

@ARTICLE{Lev25,
       author = {{Levan}, Andrew J. and {Martin-Carrillo}, Antonio and {Laskar}, Tanmoy and {Eyles-Ferris}, Rob A.~J. and {Sneppen}, Albert and {Ravasio}, Maria Edvige and {Rastinejad}, Jillian C. and {Bright}, Joe S. and {Carotenuto}, Francesco and {Chrimes}, Ashley A. and {Corcoran}, Gregory and {Gompertz}, Benjamin P. and {Jonker}, Peter G. and {Lamb}, Gavin P. and {Malesani}, Daniele B. and {Saccardi}, Andrea and {S{\'a}nchez-Sierras}, Javier and {Schneider}, Benjamin and {Schulze}, Steve and {Tanvir}, Nial R. and {Vergani}, Susanna D. and {Watson}, Darach and {An}, Jie and {Bauer}, Franz E. and {Campana}, Sergio and {Cotter}, Laura and {van Dalen}, Joyce N.~D. and {D'Elia}, Valerio and {De Pasquale}, Massimiliano and {de Ugarte Postigo}, Antonio and {Dimple} and {Hartmann}, Dieter H. and {Hjorth}, Jens and {Izzo}, Luca and {Jakobsson}, P{\'a}ll and {Kumar}, Amit and {Melandri}, Andrea and {O'Brien}, Paul and {Piranomonte}, Silvia and {Pugliese}, Giovanna and {Quirola-V{\'a}squez}, Jonathan and {Starling}, Rhaana and {Tagliaferri}, Gianpiero and {Xu}, Dong and {Wortley}, Makenzie E.},
        title = "{The Day-long, Repeating GRB 250702B: A Unique Extragalactic Transient}",
      journal = {\apjl},
         year = 2025,
        month = sep,
       volume = {990},
       number = {1},
          eid = {L28},
        pages = {L28},
          doi = {10.3847/2041-8213/adf8e1},
       adsurl = {https://ui.adsabs.harvard.edu/abs/2025ApJ...990L..28L}
}

@ARTICLE{Harr18,
       author = {{Harrison}, Richard and {Gottlieb}, Ore and {Nakar}, Ehud},
        title = "{Numerically calibrated model for propagation of a relativistic unmagnetized jet in dense media}",
      journal = {\mnras},
         year = 2018,
        month = jun,
       volume = {477},
       number = {2},
        pages = {2128-2140},
          doi = {10.1093/mnras/sty760},
archivePrefix = {arXiv},
       eprint = {1707.06234},
 primaryClass = {astro-ph.HE},
       adsurl = {https://ui.adsabs.harvard.edu/abs/2018MNRAS.477.2128H}
}

@ARTICLE{Brom11,
       author = {{Bromberg}, Omer and {Nakar}, Ehud and {Piran}, Tsvi and {Sari}, Re'em},
        title = "{The Propagation of Relativistic Jets in External Media}",
      journal = {\apj},
         year = 2011,
        month = oct,
       volume = {740},
       number = {2},
          eid = {100},
        pages = {100},
          doi = {10.1088/0004-637X/740/2/100},
archivePrefix = {arXiv},
       eprint = {1107.1326},
 primaryClass = {astro-ph.HE},
       adsurl = {https://ui.adsabs.harvard.edu/abs/2011ApJ...740..100B}
}

@ARTICLE{Brom13,
       author = {{Bromberg}, Omer and {Nakar}, Ehud and {Piran}, Tsvi and {Sari}, Re'em},
        title = "{Short versus Long and Collapsars versus Non-collapsars: A Quantitative Classification of Gamma-Ray Bursts}",
      journal = {\apj},
         year = 2013,
        month = feb,
       volume = {764},
       number = {2},
          eid = {179},
        pages = {179},
          doi = {10.1088/0004-637X/764/2/179},
archivePrefix = {arXiv},
       eprint = {1210.0068},
 primaryClass = {astro-ph.HE},
       adsurl = {https://ui.adsabs.harvard.edu/abs/2013ApJ...764..179B}
}

@ARTICLE{Kouv93,
       author = {{Kouveliotou}, Chryssa and {Meegan}, Charles A. and {Fishman}, Gerald J. and {Bhat}, Narayana P. and {Briggs}, Michael S. and {Koshut}, Thomas M. and {Paciesas}, William S. and {Pendleton}, Geoffrey N.},
        title = "{Identification of Two Classes of Gamma-Ray Bursts}",
      journal = {\apjl},
         year = 1993,
        month = aug,
       volume = {413},
        pages = {L101},
          doi = {10.1086/186969},
       adsurl = {https://ui.adsabs.harvard.edu/abs/1993ApJ...413L.101K}
}

@ARTICLE{MW99,
   author = {{MacFadyen}, A.~I. and {Woosley}, S.~E.},
    title = "{Collapsars: Gamma-Ray Bursts and Explosions in ``Failed Supernovae''}",
  journal = {\apj},
   eprint = {astro-ph/9810274},
     year = 1999,
    month = oct,
   volume = 524,
    pages = {262-289},
      doi = {10.1086/307790},
   adsurl = {https://ui.adsabs.harvard.edu/abs/1999ApJ...524..262M}
}

@misc{chakraborty2022,
  author = {Chakraborty, Angana and Dainotti, Maria and Cantrell, Olivia and Lloyd-Ronning, Nicole},
  title = {Radio-bright vs. Radio-dark Gamma-ray Bursts - More Evidence for Distinct Progenitors},
  howpublished = {arXiv:2210.12972 [astro-ph.HE]},
  year = {2022},
  note = {Draft version October 24, 2022},
  url = {https://arxiv.org/abs/2210.12972}
}

@misc{negro2024,
  author = {Negro, Michela and Cibrario, Nicoló and Burns, Eric and Wood, Joshua and Goldstein, Adam and Dal Canton, Tito},
  title = {Prompt GRB recognition through waterfalls and deep learning},
  howpublished = {arXiv:2406.03643 [astro-ph.IM]},
  year = {2024},
  note = {Draft version June 2024},
  url = {https://arxiv.org/abs/2406.03643}
}

@ARTICLE{LR23,
       author = {{Lloyd-Ronning}, Nicole and {Johnson}, Jarrett and {Cheng}, Roseanne M. and {Luu}, Ken and {Sanderbeck}, Phoebe Upton and {Kenoly}, Lailani and {Toral}, Celia},
        title = "{On the Anticorrelation between Duration and Redshift in Gamma-Ray Bursts}",
      journal = {\apj},
         year = 2023,
        month = apr,
       volume = {947},
       number = {2},
          eid = {85},
        pages = {85},
          doi = {10.3847/1538-4357/acc795},
archivePrefix = {arXiv},
       eprint = {2212.08096},
 primaryClass = {astro-ph.HE},
       adsurl = {https://ui.adsabs.harvard.edu/abs/2023ApJ...947...85L}
}

@article{LR22,
  author = {Lloyd-Ronning, Nicole},
  title = {Radio-loud versus Radio-quiet Gamma-Ray Bursts: The Role of Binary Progenitors},
  journal = {The Astrophysical Journal},
  volume = {928},
  number = {2},
  pages = {104},
  year = {2022},
  doi = {10.3847/1538-4357/ac54b3},
  url = {https://doi.org/10.3847/1538-4357/ac54b3}
}

@ARTICLE{LR19b,
   author = {{Lloyd-Ronning}, N.~M. and {Aykutalp}, A. and {Johnson}, J.~L.
	},
    title = "{On the cosmological evolution of long gamma-ray burst properties}",
  journal = {\mnras},
archivePrefix = "arXiv",
   eprint = {1906.02278},
 primaryClass = "astro-ph.HE",
     year = 2019,
    month = oct,
   volume = 488,
    pages = {5823-5832},
      doi = {10.1093/mnras/stz2155},
   adsurl = {https://ui.adsabs.harvard.edu/abs/2019MNRAS.488.5823L}
}

@ARTICLE{LR24,
       author = {{Lloyd-Ronning}, Nicole M. and {Johnson}, Jarrett and {Upton Sanderbeck}, Phoebe and {Silva}, Makana and {Cheng}, Roseanne M.},
        title = "{White dwarf-black hole binary progenitors of low-redshift gamma-ray bursts}",
      journal = {\mnras},
         year = 2024,
        month = dec,
       volume = {535},
       number = {3},
        pages = {2800-2811},
          doi = {10.1093/mnras/stae2502},
archivePrefix = {arXiv},
       eprint = {2408.12654},
 primaryClass = {astro-ph.HE},
       adsurl = {https://ui.adsabs.harvard.edu/abs/2024MNRAS.535.2800L}
}

@ARTICLE{LR20b,
       author = {{Lloyd-Ronning}, Nicole M. and {Johnson}, Jarrett L. and {Aykutalp}, Aycin},
        title = "{The consequences of gamma-ray burst jet opening angle evolution on the inferred star formation rate}",
      journal = {\mnras},
         year = 2020,
        month = nov,
       volume = {498},
       number = {4},
        pages = {5041-5047},
          doi = {10.1093/mnras/staa2787},
archivePrefix = {arXiv},
       eprint = {2006.00022},
 primaryClass = {astro-ph.HE},
       adsurl = {https://ui.adsabs.harvard.edu/abs/2020MNRAS.498.5041L}
}

@ARTICLE{Pet15,
       author = {{Petrosian}, Vah{\'e} and {Kitanidis}, Ellie and {Kocevski}, Daniel},
        title = "{Cosmological Evolution of Long Gamma-Ray Bursts and the Star Formation Rate}",
      journal = {\apj},
         year = 2015,
        month = jun,
       volume = {806},
       number = {1},
          eid = {44},
        pages = {44},
          doi = {10.1088/0004-637X/806/1/44},
archivePrefix = {arXiv},
       eprint = {1504.01414},
 primaryClass = {astro-ph.HE},
       adsurl = {https://ui.adsabs.harvard.edu/abs/2015ApJ...806...44P}
}

@ARTICLE{EP92,
   author = {{Efron}, B. and {Petrosian}, V.},
    title = "{A simple test of independence for truncated data with applications to redshift surveys}",
  journal = {\apj},
     year = 1992,
    month = nov,
   volume = 399,
    pages = {345-352},
      doi = {10.1086/171931},
   adsurl = {http://adsabs.harvard.edu/abs/1992ApJ...399..345E}
}

@ARTICLE{LB71,
   author = {{Lynden-Bell}, D.},
    title = "{A method of allowing for known observational selection in small samples applied to 3CR quasars}",
  journal = {\mnras},
     year = 1971,
   volume = 155,
    pages = {95},
      doi = {10.1093/mnras/155.1.95},
   adsurl = {http://adsabs.harvard.edu/abs/1971MNRAS.155...95L}
}

@ARTICLE{WP15,
       author = {{Wanderman}, David and {Piran}, Tsvi},
        title = "{The rate, luminosity function and time delay of non-Collapsar short GRBs}",
      journal = {\mnras},
         year = 2015,
        month = apr,
       volume = {448},
       number = {4},
        pages = {3026-3037},
          doi = {10.1093/mnras/stv123},
archivePrefix = {arXiv},
       eprint = {1405.5878},
 primaryClass = {astro-ph.HE},
       adsurl = {https://ui.adsabs.harvard.edu/abs/2015MNRAS.448.3026W}
}

@ARTICLE{Bloom02,
       author = {{Bloom}, J.~S. and {Kulkarni}, S.~R. and {Djorgovski}, S.~G.},
        title = "{The Observed Offset Distribution of Gamma-Ray Bursts from Their Host Galaxies: A Robust Clue to the Nature of the Progenitors}",
      journal = {\aj},
         year = 2002,
        month = mar,
       volume = {123},
       number = {3},
        pages = {1111-1148},
          doi = {10.1086/338893},
archivePrefix = {arXiv},
       eprint = {astro-ph/0010176},
 primaryClass = {astro-ph},
       adsurl = {https://ui.adsabs.harvard.edu/abs/2002AJ....123.1111B}
}

@ARTICLE{Ly17,
       author = {{Lyman}, J.~D. and {Levan}, A.~J. and {Tanvir}, N.~R. and {Fynbo}, J.~P.~U. and {McGuire}, J.~T.~W. and {Perley}, D.~A. and {Angus}, C.~R. and {Bloom}, J.~S. and {Conselice}, C.~J. and {Fruchter}, A.~S. and {Hjorth}, J. and {Jakobsson}, P. and {Starling}, R.~L.~C.},
        title = "{The host galaxies and explosion sites of long-duration gamma ray bursts: Hubble Space Telescope near-infrared imaging}",
      journal = {\mnras},
         year = 2017,
        month = may,
       volume = {467},
       number = {2},
        pages = {1795-1817},
          doi = {10.1093/mnras/stx220},
archivePrefix = {arXiv},
       eprint = {1701.05925},
 primaryClass = {astro-ph.GA},
       adsurl = {https://ui.adsabs.harvard.edu/abs/2017MNRAS.467.1795L}
}

@ARTICLE{Yu15,
       author = {{Yu}, H. and {Wang}, F.~Y. and {Dai}, Z.~G. and {Cheng}, K.~S.},
        title = "{An Unexpectedly Low-redshift Excess of Swift Gamma-ray Burst Rate}",
      journal = {\apjs},
         year = 2015,
        month = may,
       volume = {218},
       number = {1},
          eid = {13},
        pages = {13},
          doi = {10.1088/0067-0049/218/1/13},
archivePrefix = {arXiv},
       eprint = {1504.01812},
 primaryClass = {astro-ph.HE},
       adsurl = {https://ui.adsabs.harvard.edu/abs/2015ApJS..218...13Y}
}

@ARTICLE{Tsv17,
       author = {{Tsvetkova}, A. and {Frederiks}, D. and {Golenetskii}, S. and {Lysenko}, A. and {Oleynik}, P. and {Pal'shin}, V. and {Svinkin}, D. and {Ulanov}, M. and {Cline}, T. and {Hurley}, K. and {Aptekar}, R.},
        title = "{The Konus-Wind Catalog of Gamma-Ray Bursts with Known Redshifts. I. Bursts Detected in the Triggered Mode}",
      journal = {\apj},
         year = 2017,
        month = dec,
       volume = {850},
       number = {2},
          eid = {161},
        pages = {161},
          doi = {10.3847/1538-4357/aa96af},
archivePrefix = {arXiv},
       eprint = {1710.08746},
 primaryClass = {astro-ph.HE},
       adsurl = {https://ui.adsabs.harvard.edu/abs/2017ApJ...850..161T}
}

@ARTICLE{Vir09,
       author = {{Virgili}, Francisco J. and {Liang}, En-Wei and {Zhang}, Bing},
        title = "{Low-luminosity gamma-ray bursts as a distinct GRB population: a firmer case from multiple criteria constraints}",
      journal = {\mnras},
         year = 2009,
        month = jan,
       volume = {392},
       number = {1},
        pages = {91-103},
          doi = {10.1111/j.1365-2966.2008.14063.x},
archivePrefix = {arXiv},
       eprint = {0801.4751},
 primaryClass = {astro-ph},
       adsurl = {https://ui.adsabs.harvard.edu/abs/2009MNRAS.392...91V}
}

@ARTICLE{Bonn97,
       author = {{Bonnell}, J.~T. and {Norris}, J.~P. and {Nemiroff}, R.~J. and {Scargle}, J.~D.},
        title = "{Brightness-independent Measurements of Gamma-Ray Burst Durations}",
      journal = {\apj},
         year = 1997,
        month = nov,
       volume = {490},
       number = {1},
        pages = {79-91},
          doi = {10.1086/304841},
       adsurl = {https://ui.adsabs.harvard.edu/abs/1997ApJ...490...79B}
}

@ARTICLE{Per16,
       author = {{Perley}, D.~A. and {Kr{\"u}hler}, T. and {Schulze}, S. and {de Ugarte Postigo}, A. and {Hjorth}, J. and {Berger}, E. and {Cenko}, S.~B. and {Chary}, R. and {Cucchiara}, A. and {Ellis}, R. and {Fong}, W. and {Fynbo}, J.~P.~U. and {Gorosabel}, J. and {Greiner}, J. and {Jakobsson}, P. and {Kim}, S. and {Laskar}, T. and {Levan}, A.~J. and {Micha{\l}owski}, M.~J. and {Milvang-Jensen}, B. and {Tanvir}, N.~R. and {Th{\"o}ne}, C.~C. and {Wiersema}, K.},
        title = "{The Swift Gamma-Ray Burst Host Galaxy Legacy Survey. I. Sample Selection and Redshift Distribution}",
      journal = {\apj},
         year = 2016,
        month = jan,
       volume = {817},
       number = {1},
          eid = {7},
        pages = {7},
          doi = {10.3847/0004-637X/817/1/7},
archivePrefix = {arXiv},
       eprint = {1504.02482},
 primaryClass = {astro-ph.GA},
       adsurl = {https://ui.adsabs.harvard.edu/abs/2016ApJ...817....7P}
}

@ARTICLE{LM17,
       author = {{Le}, Truong and {Mehta}, Vedant},
        title = "{Revisiting the Redshift Distribution of Gamma-Ray Bursts in the Swift Era}",
      journal = {\apj},
         year = 2017,
        month = mar,
       volume = {837},
       number = {1},
          eid = {17},
        pages = {17},
          doi = {10.3847/1538-4357/aa5fa7},
archivePrefix = {arXiv},
       eprint = {1702.03338},
 primaryClass = {astro-ph.HE},
       adsurl = {https://ui.adsabs.harvard.edu/abs/2017ApJ...837...17L}
}

@ARTICLE{Dai21,
       author = {{Dainotti}, M.~G. and {Petrosian}, V. and {Bowden}, L.},
        title = "{Cosmological Evolution of the Formation Rate of Short Gamma-Ray Bursts with and without Extended Emission}",
      journal = {\apjl},
         year = 2021,
        month = jun,
       volume = {914},
       number = {2},
          eid = {L40},
        pages = {L40},
          doi = {10.3847/2041-8213/abf5e4},
archivePrefix = {arXiv},
       eprint = {2104.13555},
 primaryClass = {astro-ph.HE},
       adsurl = {https://ui.adsabs.harvard.edu/abs/2021ApJ...914L..40D}
}

@ARTICLE{Hal23,
       author = {{Halevi}, Goni and {Wu}, Belinda and {M{\"o}sta}, Philipp and {Gottlieb}, Ore and {Tchekhovskoy}, Alexander and {Aguilera-Dena}, David R.},
        title = "{Density Profiles of Collapsed Rotating Massive Stars Favor Long Gamma-Ray Bursts}",
      journal = {\apjl},
         year = 2023,
        month = feb,
       volume = {944},
       number = {2},
          eid = {L38},
        pages = {L38},
          doi = {10.3847/2041-8213/acb702},
archivePrefix = {arXiv},
       eprint = {2211.11781},
 primaryClass = {astro-ph.HE},
       adsurl = {https://ui.adsabs.harvard.edu/abs/2023ApJ...944L..38H}
}

@ARTICLE{Brom16,
       author = {{Bromberg}, Omer and {Tchekhovskoy}, Alexander},
        title = "{Relativistic MHD simulations of core-collapse GRB jets: 3D instabilities and magnetic dissipation}",
      journal = {\mnras},
         year = 2016,
        month = feb,
       volume = {456},
       number = {2},
        pages = {1739-1760},
          doi = {10.1093/mnras/stv2591},
archivePrefix = {arXiv},
       eprint = {1508.02721},
 primaryClass = {astro-ph.HE},
       adsurl = {https://ui.adsabs.harvard.edu/abs/2016MNRAS.456.1739B}
}

@ARTICLE{AgDen20,
       author = {{Aguilera-Dena}, David R. and {Langer}, Norbert and {Antoniadis}, John and {M{\"u}ller}, Bernhard},
        title = "{Precollapse Properties of Superluminous Supernovae and Long Gamma-Ray Burst Progenitor Models}",
      journal = {\apj},
         year = 2020,
        month = oct,
       volume = {901},
       number = {2},
          eid = {114},
        pages = {114},
          doi = {10.3847/1538-4357/abb138},
archivePrefix = {arXiv},
       eprint = {2008.09132},
 primaryClass = {astro-ph.SR},
       adsurl = {https://ui.adsabs.harvard.edu/abs/2020ApJ...901..114A}
}

@ARTICLE{Matz03,
       author = {{Matzner}, Christopher D.},
        title = "{Supernova hosts for gamma-ray burst jets: dynamical constraints}",
      journal = {\mnras},
         year = 2003,
        month = oct,
       volume = {345},
       number = {2},
        pages = {575-589},
          doi = {10.1046/j.1365-8711.2003.06969.x},
archivePrefix = {arXiv},
       eprint = {astro-ph/0203085},
 primaryClass = {astro-ph},
       adsurl = {https://ui.adsabs.harvard.edu/abs/2003MNRAS.345..575M}
}

@ARTICLE{Le20,
       author = {{Le}, Truong and {Ratke}, Cecilia and {Mehta}, Vedant},
        title = "{Resolving the excess of long GRB's at low redshift in the Swift era}",
      journal = {\mnras},
         year = 2020,
        month = mar,
       volume = {493},
       number = {1},
        pages = {1479-1491},
          doi = {10.1093/mnras/staa366},
archivePrefix = {arXiv},
       eprint = {2002.02110},
 primaryClass = {astro-ph.HE},
       adsurl = {https://ui.adsabs.harvard.edu/abs/2020MNRAS.493.1479L}
}

@ARTICLE{Pet24,
       author = {{Petrosian}, Vah{\'e} and {Dainotti}, Maria G.},
        title = "{Progenitors of Low-redshift Gamma-Ray Bursts}",
      journal = {\apjl},
         year = 2024,
        month = mar,
       volume = {963},
       number = {1},
          eid = {L12},
        pages = {L12},
          doi = {10.3847/2041-8213/ad2763},
archivePrefix = {arXiv},
       eprint = {2305.15081},
 primaryClass = {astro-ph.HE},
       adsurl = {https://ui.adsabs.harvard.edu/abs/2024ApJ...963L..12P}
}

@ARTICLE{Gal98,
       author = {{Galama}, T.~J. and {Vreeswijk}, P.~M. and {van Paradijs}, J. and
         {Kouveliotou}, C. and {Augusteijn}, T. and {B{\"o}hnhardt}, H. and
         {Brewer}, J.~P. and {Doublier}, V. and {Gonzalez}, J. -F. and
         {Leibundgut}, B. and {Lidman}, C. and {Hainaut}, O.~R. and {Patat}, F. and
         {Heise}, J. and {in't Zand}, J. and {Hurley}, K. and {Groot}, P.~J. and
         {Strom}, R.~G. and {Mazzali}, P.~A. and {Iwamoto}, K. and {Nomoto}, K. and
         {Umeda}, H. and {Nakamura}, T. and {Young}, T.~R. and {Suzuki}, T. and
         {Shigeyama}, T. and {Koshut}, T. and {Kippen}, M. and {Robinson}, C. and
         {de Wildt}, P. and {Wijers}, R.~A.~M.~J. and {Tanvir}, N. and
         {Greiner}, J. and {Pian}, E. and {Palazzi}, E. and {Frontera}, F. and
         {Masetti}, N. and {Nicastro}, L. and {Feroci}, M. and {Costa}, E. and
         {Piro}, L. and {Peterson}, B.~A. and {Tinney}, C. and {Boyle}, B. and
         {Cannon}, R. and {Stathakis}, R. and {Sadler}, E. and {Begam}, M.~C. and
         {Ianna}, P.},
        title = "{An unusual supernova in the error box of the {\ensuremath{\gamma}}-ray burst of 25 April 1998}",
      journal = {\nat},
         year = "1998",
        month = "Oct",
       volume = {395},
       number = {6703},
        pages = {670-672},
          doi = {10.1038/27150},
archivePrefix = {arXiv},
       eprint = {astro-ph/9806175},
 primaryClass = {astro-ph},
       adsurl = {https://ui.adsabs.harvard.edu/abs/1998Natur.395..670G}
}

@ARTICLE{Hjorth03,
   author = {{Hjorth}, J. and {Sollerman}, J. and {M{\o}ller}, P. and {Fynbo}, J.~P.~U. and 
	{Woosley}, S.~E. and {Kouveliotou}, C. and {Tanvir}, N.~R. and 
	{Greiner}, J. and {Andersen}, M.~I. and {Castro-Tirado}, A.~J. and 
	{Castro Cer{\'o}n}, J.~M. and {Fruchter}, A.~S. and {Gorosabel}, J. and 
	{Jakobsson}, P. and {Kaper}, L. and {Klose}, S. and {Masetti}, N. and 
	{Pedersen}, H. and {Pedersen}, K. and {Pian}, E. and {Palazzi}, E. and 
	{Rhoads}, J.~E. and {Rol}, E. and {van den Heuvel}, E.~P.~J. and 
	{Vreeswijk}, P.~M. and {Watson}, D. and {Wijers}, R.~A.~M.~J.
	},
    title = "{A very energetic supernova associated with the {$\gamma$}-ray burst of 29 March 2003}",
  journal = {\nat},
   eprint = {astro-ph/0306347},
     year = 2003,
    month = jun,
   volume = 423,
    pages = {847-850},
      doi = {10.1038/nature01750},
   adsurl = {http://adsabs.harvard.edu/abs/2003Natur.423..847H}
}

@article{HB12,
  title={The GRB-supernova connection},
  author={Hjorth, Jens and Bloom, Joshua S},
  journal={Gamma-ray bursts},
  year={2012},
  publisher={Cambridge University Press, Cambridge}
}

@ARTICLE{Chen24,
       author = {{Chen}, Junping and {Shen}, Rong-Feng and {Tan}, Wen-Jun and {Wang}, Chen-Wei and {Xiong}, Shao-Lin and {Chen}, Run-Chao and {Zhang}, Bin-Bin},
        title = "{Repeated partial disruptions in a WD-NS or WD-BH merger modulate the prompt emission of long-duration merger-type GRBs}",
      journal = {arXiv e-prints},
         year = 2024,
        month = aug,
          eid = {arXiv:2409.00472},
        pages = {arXiv:2409.00472},
archivePrefix = {arXiv},
       eprint = {2409.00472},
 primaryClass = {astro-ph.HE},
       adsurl = {https://ui.adsabs.harvard.edu/abs/2024arXiv240900472C}
}

@ARTICLE{Zhu22,
       author = {{Zhu}, Jin-Ping and {Wang}, Xiangyu Ivy and {Sun}, Hui and {Yang}, Yuan-Pei and {Li}, Zhuo and {Hu}, Rui-Chong and {Qin}, Ying and {Wu}, Shichao},
        title = "{Long-duration Gamma-Ray Burst and Associated Kilonova Emission from Fast-spinning Black Hole-Neutron Star Mergers}",
      journal = {\apjl},
         year = 2022,
        month = sep,
       volume = {936},
       number = {1},
          eid = {L10},
        pages = {L10},
          doi = {10.3847/2041-8213/ac85ad},
archivePrefix = {arXiv},
       eprint = {2207.10470},
 primaryClass = {astro-ph.HE},
       adsurl = {https://ui.adsabs.harvard.edu/abs/2022ApJ...936L..10Z}
}

@ARTICLE{Chrimes25,
       author = {{Chrimes}, A.~A. and {Gaspari}, N. and {Levan}, A.~J. and {Briel}, M.~M. and {Eldridge}, J.~J. and {Gompertz}, B.~P. and {Nelemans}, G. and {Nugent}, A.~E. and {Rastinejad}, J.~C. and {van Zeist}, W.~G.~J.},
        title = "{White dwarf-neutron star binaries: a plausible pathway for long-duration gamma-ray bursts from compact object mergers?}",
      journal = {arXiv e-prints},
         year = 2025,
        month = aug,
          eid = {arXiv:2508.10984},
        pages = {arXiv:2508.10984},
          doi = {10.48550/arXiv.2508.10984},
archivePrefix = {arXiv},
       eprint = {2508.10984},
 primaryClass = {astro-ph.HE},
       adsurl = {https://ui.adsabs.harvard.edu/abs/2025arXiv250810984C}
}

@ARTICLE{Nik25,
       author = {{Khatiya}, Nikita S. and {Dainotti}, Maria Giovanna and {Narendra}, Aditya and {Bal}, Dhruv S. and {Lenart}, Aleksander {\L}. and {Hartmann}, Dieter H.},
        title = "{Probing Evolution of Long Gamma-Ray Burst Properties through Their Cosmic Formation History}",
      journal = {\apj},
         year = 2025,
        month = sep,
       volume = {990},
       number = {1},
          eid = {69},
        pages = {69},
          doi = {10.3847/1538-4357/adf219},
archivePrefix = {arXiv},
       eprint = {2508.20093},
 primaryClass = {astro-ph.HE},
       adsurl = {https://ui.adsabs.harvard.edu/abs/2025ApJ...990...69K}
}

@article{WB06,
  title={The supernova--gamma-ray burst connection},
  author={Woosley, SE and Bloom, JS},
  journal={Annu. Rev. Astron. Astrophys.},
  volume={44},
  pages={507--556},
  year={2006},
  publisher={Annual Reviews}
}

@ARTICLE{Nak06,
       author = {{Nakar}, Ehud and {Gal-Yam}, Avishay and {Fox}, Derek B.},
        title = "{The Local Rate and the Progenitor Lifetimes of Short-Hard Gamma-Ray Bursts: Synthesis and Predictions for the Laser Interferometer Gravitational-Wave Observatory}",
      journal = {\apj},
         year = 2006,
        month = oct,
       volume = {650},
       number = {1},
        pages = {281-290},
          doi = {10.1086/505855},
archivePrefix = {arXiv},
       eprint = {astro-ph/0511254},
 primaryClass = {astro-ph},
       adsurl = {https://ui.adsabs.harvard.edu/abs/2006ApJ...650..281N}
}

@ARTICLE{Berg07,
       author = {{Berger}, E. and {Fox}, D.~B. and {Price}, P.~A. and {Nakar}, E. and {Gal-Yam}, A. and {Holz}, D.~E. and {Schmidt}, B.~P. and {Cucchiara}, A. and {Cenko}, S.~B. and {Kulkarni}, S.~R. and {Soderberg}, A.~M. and {Frail}, D.~A. and {Penprase}, B.~E. and {Rau}, A. and {Ofek}, E. and {Burnell}, S.~J. Bell and {Cameron}, P.~B. and {Cowie}, L.~L. and {Dopita}, M.~A. and {Hook}, I. and {Peterson}, B.~A. and {Podsiadlowski}, P. and {Roth}, K.~C. and {Rutledge}, R.~E. and {Sheppard}, S.~S. and {Songaila}, A.},
        title = "{A New Population of High-Redshift Short-Duration Gamma-Ray Bursts}",
      journal = {\apj},
         year = 2007,
        month = aug,
       volume = {664},
       number = {2},
        pages = {1000-1010},
          doi = {10.1086/518762},
archivePrefix = {arXiv},
       eprint = {astro-ph/0611128},
 primaryClass = {astro-ph},
       adsurl = {https://ui.adsabs.harvard.edu/abs/2007ApJ...664.1000B}
}

@ARTICLE{Pir92,
       author = {{Piran}, Tsvi},
        title = "{The Implications of the Compton (GRO) Observations for Cosmological Gamma-Ray Bursts}",
      journal = {\apjl},
         year = 1992,
        month = apr,
       volume = {389},
        pages = {L45},
          doi = {10.1086/186345},
       adsurl = {https://ui.adsabs.harvard.edu/abs/1992ApJ...389L..45P}
}

@ARTICLE{Cow12,
       author = {{Coward}, D.~M. and {Howell}, E.~J. and {Piran}, T. and {Stratta}, G. and {Branchesi}, M. and {Bromberg}, O. and {Gendre}, B. and {Burman}, R.~R. and {Guetta}, D.},
        title = "{The Swift short gamma-ray burst rate density: implications for binary neutron star merger rates}",
      journal = {\mnras},
         year = 2012,
        month = oct,
       volume = {425},
       number = {4},
        pages = {2668-2673},
          doi = {10.1111/j.1365-2966.2012.21604.x},
archivePrefix = {arXiv},
       eprint = {1202.2179},
 primaryClass = {astro-ph.CO},
       adsurl = {https://ui.adsabs.harvard.edu/abs/2012MNRAS.425.2668C}
}

@ARTICLE{GP06,
       author = {{Guetta}, D. and {Piran}, T.},
        title = "{The BATSE-Swift luminosity and redshift distributions of short-duration GRBs}",
      journal = {\aap},
         year = 2006,
        month = jul,
       volume = {453},
       number = {3},
        pages = {823-828},
          doi = {10.1051/0004-6361:20054498},
archivePrefix = {arXiv},
       eprint = {astro-ph/0511239},
 primaryClass = {astro-ph},
       adsurl = {https://ui.adsabs.harvard.edu/abs/2006A&A...453..823G}
}

@ARTICLE{Ana18,
       author = {{Anand}, Nikhil and {Shahid}, Mustafa and {Resmi}, Lekshmi},
        title = "{Merger delay time distribution of extended emission short GRBs}",
      journal = {\mnras},
         year = 2018,
        month = dec,
       volume = {481},
       number = {4},
        pages = {4332-4341},
          doi = {10.1093/mnras/sty2530},
archivePrefix = {arXiv},
       eprint = {1710.04996},
 primaryClass = {astro-ph.HE},
       adsurl = {https://ui.adsabs.harvard.edu/abs/2018MNRAS.481.4332A}
}

@ARTICLE{Bel18,
       author = {{Belczynski}, K. and {Bulik}, T. and {Olejak}, A. and {Chruslinska}, M. and {Singh}, N. and {Pol}, N. and {Zdunik}, L. and {O'Shaughnessy}, R. and {McLaughlin}, M. and {Lorimer}, D. and {Korobkin}, O. and {van den Heuvel}, E.~P.~J. and {Davies}, M.~B. and {Holz}, D.~E.},
        title = "{Binary neutron star formation and the origin of GW170817}",
      journal = {arXiv e-prints},
         year = 2018,
        month = dec,
          eid = {arXiv:1812.10065},
        pages = {arXiv:1812.10065},
          doi = {10.48550/arXiv.1812.10065},
archivePrefix = {arXiv},
       eprint = {1812.10065},
 primaryClass = {astro-ph.HE},
       adsurl = {https://ui.adsabs.harvard.edu/abs/2018arXiv181210065B}
}

@ARTICLE{Broe22,
       author = {{Broekgaarden}, Floor S. and {Berger}, Edo and {Stevenson}, Simon and {Justham}, Stephen and {Mandel}, Ilya and {Chru{\'s}li{\'n}ska}, Martyna and {van Son}, Lieke A.~C. and {Wagg}, Tom and {Vigna-G{\'o}mez}, Alejandro and {de Mink}, Selma E. and {Chattopadhyay}, Debatri and {Neijssel}, Coenraad J.},
        title = "{Impact of massive binary star and cosmic evolution on gravitational wave observations - II. Double compact object rates and properties}",
      journal = {\mnras},
         year = 2022,
        month = nov,
       volume = {516},
       number = {4},
        pages = {5737-5761},
          doi = {10.1093/mnras/stac1677},
archivePrefix = {arXiv},
       eprint = {2112.05763},
 primaryClass = {astro-ph.HE},
       adsurl = {https://ui.adsabs.harvard.edu/abs/2022MNRAS.516.5737B}
}

@ARTICLE{Chr18,
       author = {{Chruslinska}, Martyna and {Belczynski}, Krzysztof and {Klencki}, Jakub and {Benacquista}, Matthew},
        title = "{Double neutron stars: merger rates revisited}",
      journal = {\mnras},
         year = 2018,
        month = mar,
       volume = {474},
       number = {3},
        pages = {2937-2958},
          doi = {10.1093/mnras/stx2923},
archivePrefix = {arXiv},
       eprint = {1708.07885},
 primaryClass = {astro-ph.HE},
       adsurl = {https://ui.adsabs.harvard.edu/abs/2018MNRAS.474.2937C}
}

@ARTICLE{Santo22,
       author = {{Santoliquido}, Filippo and {Mapelli}, Michela and {Artale}, M. Celeste and {Boco}, Lumen},
        title = "{Modelling the host galaxies of binary compact object mergers with observational scaling relations}",
      journal = {\mnras},
         year = 2022,
        month = nov,
       volume = {516},
       number = {3},
        pages = {3297-3317},
          doi = {10.1093/mnras/stac2384},
archivePrefix = {arXiv},
       eprint = {2205.05099},
 primaryClass = {astro-ph.HE},
       adsurl = {https://ui.adsabs.harvard.edu/abs/2022MNRAS.516.3297S}
}

@ARTICLE{Zev22,
       author = {{Zevin}, Michael and {Nugent}, Anya E. and {Adhikari}, Susmita and {Fong}, Wen-fai and {Holz}, Daniel E. and {Kelley}, Luke Zoltan},
        title = "{Observational Inference on the Delay Time Distribution of Short Gamma-Ray Bursts}",
      journal = {\apjl},
         year = 2022,
        month = nov,
       volume = {940},
       number = {1},
          eid = {L18},
        pages = {L18},
          doi = {10.3847/2041-8213/ac91cd},
archivePrefix = {arXiv},
       eprint = {2206.02814},
 primaryClass = {astro-ph.HE},
       adsurl = {https://ui.adsabs.harvard.edu/abs/2022ApJ...940L..18Z}
}

@ARTICLE{Has24,
       author = {{Hasan}, Ali M. and {Azzam}, Walid J.},
        title = "{Does the Redshift Distribution of Swift Long GRBs Trace the Star-Formation Rate?}",
      journal = {International Journal of Astronomy and Astrophysics},
         year = 2024,
        month = jan,
       volume = {14},
       number = {1},
        pages = {20-44},
          doi = {10.4236/ijaa.2024.141002},
archivePrefix = {arXiv},
       eprint = {2403.15087},
 primaryClass = {astro-ph.HE},
       adsurl = {https://ui.adsabs.harvard.edu/abs/2024IJAA...14...20H}
}

@ARTICLE{Fry99,
       author = {{Fryer}, Chris L. and {Woosley}, S.~E. and {Herant}, Marc and {Davies}, Melvyn B.},
        title = "{Merging White Dwarf/Black Hole Binaries and Gamma-Ray Bursts}",
      journal = {\apj},
         year = 1999,
        month = aug,
       volume = {520},
       number = {2},
        pages = {650-660},
          doi = {10.1086/307467},
archivePrefix = {arXiv},
       eprint = {astro-ph/9808094},
 primaryClass = {astro-ph},
       adsurl = {https://ui.adsabs.harvard.edu/abs/1999ApJ...520..650F}
}

@ARTICLE{Rast22,
       author = {{Rastinejad}, Jillian C. and {Gompertz}, Benjamin P. and {Levan}, Andrew J. and {Fong}, Wen-fai and {Nicholl}, Matt and {Lamb}, Gavin P. and {Malesani}, Daniele B. and {Nugent}, Anya E. and {Oates}, Samantha R. and {Tanvir}, Nial R. and {de Ugarte Postigo}, Antonio and {Kilpatrick}, Charles D. and {Moore}, Christopher J. and {Metzger}, Brian D. and {Ravasio}, Maria Edvige and {Rossi}, Andrea and {Schroeder}, Genevieve and {Jencson}, Jacob and {Sand}, David J. and {Smith}, Nathan and {Ag{\"u}{\'\i} Fern{\'a}ndez}, Jos{\'e} Feliciano and {Berger}, Edo and {Blanchard}, Peter K. and {Chornock}, Ryan and {Cobb}, Bethany E. and {De Pasquale}, Massimiliano and {Fynbo}, Johan P.~U. and {Izzo}, Luca and {Kann}, D. Alexander and {Laskar}, Tanmoy and {Marini}, Ester and {Paterson}, Kerry and {Escorial}, Alicia Rouco and {Sears}, Huei M. and {Th{\"o}ne}, Christina C.},
        title = "{A kilonova following a long-duration gamma-ray burst at 350 Mpc}",
      journal = {\nat},
         year = 2022,
        month = dec,
       volume = {612},
       number = {7939},
        pages = {223-227},
          doi = {10.1038/s41586-022-05390-w},
archivePrefix = {arXiv},
       eprint = {2204.10864},
 primaryClass = {astro-ph.HE},
       adsurl = {https://ui.adsabs.harvard.edu/abs/2022Natur.612..223R}
}

@ARTICLE{Zhang2022,
       author = {{Zhang}, Hai-Ming and {Huang}, Yi-Yun and {Zheng}, Jian-He and {Liu}, Ruo-Yu and {Wang}, Xiang-Yu},
        title = "{Fermi-LAT Detection of a GeV Afterglow from a Compact Stellar Merger}",
      journal = {\apjl},
         year = 2022,
        month = jul,
       volume = {933},
       number = {1},
          eid = {L22},
        pages = {L22},
          doi = {10.3847/2041-8213/ac7b23},
archivePrefix = {arXiv},
       eprint = {2205.09675},
 primaryClass = {astro-ph.HE},
       adsurl = {https://ui.adsabs.harvard.edu/abs/2022ApJ...933L..22Z}
}

@ARTICLE{Troja22,
       author = {{Troja}, E. and {Fryer}, C.~L. and {O'Connor}, B. and {Ryan}, G. and {Dichiara}, S. and {Kumar}, A. and {Ito}, N. and {Gupta}, R. and {Wollaeger}, R.~T. and {Norris}, J.~P. and {Kawai}, N. and {Butler}, N.~R. and {Aryan}, A. and {Misra}, K. and {Hosokawa}, R. and {Murata}, K.~L. and {Niwano}, M. and {Pandey}, S.~B. and {Kutyrev}, A. and {van Eerten}, H.~J. and {Chase}, E.~A. and {Hu}, Y. -D. and {Caballero-Garcia}, M.~D. and {Castro-Tirado}, A.~J.},
        title = "{A nearby long gamma-ray burst from a merger of compact objects}",
      journal = {\nat},
         year = 2022,
        month = dec,
       volume = {612},
       number = {7939},
        pages = {228-231},
          doi = {10.1038/s41586-022-05327-3},
archivePrefix = {arXiv},
       eprint = {2209.03363},
 primaryClass = {astro-ph.HE},
       adsurl = {https://ui.adsabs.harvard.edu/abs/2022Natur.612..228T}
}

@ARTICLE{Ghir15,
       author = {{Ghirlanda}, G. and {Bernardini}, M.~G. and {Calderone}, G. and {D'Avanzo}, P.},
        title = "{Are short Gamma Ray Bursts similar to long ones?}",
      journal = {Journal of High Energy Astrophysics},
         year = 2015,
        month = sep,
       volume = {7},
        pages = {81-89},
          doi = {10.1016/j.jheap.2015.04.002},
       adsurl = {https://ui.adsabs.harvard.edu/abs/2015JHEAp...7...81G}
}

@ARTICLE{Cast25,
       author = {{Castrejon}, Cristian and {Nugent}, Anya E. and {Fong}, Wen-fai and {Schroeder}, Genevieve and {Rouco Escorial}, Alicia and {Guerra}, Olivia},
        title = "{Exploring the Relationship Between Swift Short Gamma-Ray Burst Afterglows and their Host Galaxy Properties}",
      journal = {arXiv e-prints},
         year = 2025,
        month = aug,
          eid = {arXiv:2508.20156},
        pages = {arXiv:2508.20156},
          doi = {10.48550/arXiv.2508.20156},
archivePrefix = {arXiv},
       eprint = {2508.20156},
 primaryClass = {astro-ph.HE},
       adsurl = {https://ui.adsabs.harvard.edu/abs/2025arXiv250820156C}
}

@ARTICLE{Ghir16,
       author = {{Ghirlanda}, G. and {Salafia}, O.~S. and {Pescalli}, A. and {Ghisellini}, G. and {Salvaterra}, R. and {Chassande-Mottin}, E. and {Colpi}, M. and {Nappo}, F. and {D'Avanzo}, P. and {Melandri}, A. and {Bernardini}, M.~G. and {Branchesi}, M. and {Campana}, S. and {Ciolfi}, R. and {Covino}, S. and {G{\"o}tz}, D. and {Vergani}, S.~D. and {Zennaro}, M. and {Tagliaferri}, G.},
        title = "{Short gamma-ray bursts at the dawn of the gravitational wave era}",
      journal = {\aap},
         year = 2016,
        month = oct,
       volume = {594},
          eid = {A84},
        pages = {A84},
          doi = {10.1051/0004-6361/201628993},
archivePrefix = {arXiv},
       eprint = {1607.07875},
 primaryClass = {astro-ph.HE},
       adsurl = {https://ui.adsabs.harvard.edu/abs/2016A&A...594A..84G}
}

@ARTICLE{Dav15,
       author = {{D'Avanzo}, P.},
        title = "{Short gamma-ray bursts: A review}",
      journal = {Journal of High Energy Astrophysics},
         year = 2015,
        month = sep,
       volume = {7},
        pages = {73-80},
          doi = {10.1016/j.jheap.2015.07.002},
       adsurl = {https://ui.adsabs.harvard.edu/abs/2015JHEAp...7...73D}
}

@ARTICLE{Levan24,
       author = {{Levan}, Andrew J. and {Gompertz}, Benjamin P. and {Salafia}, Om Sharan and {Bulla}, Mattia and {Burns}, Eric and {Hotokezaka}, Kenta and {Izzo}, Luca and {Lamb}, Gavin P. and {Malesani}, Daniele B. and {Oates}, Samantha R. and {Ravasio}, Maria Edvige and {Rouco Escorial}, Alicia and {Schneider}, Benjamin and {Sarin}, Nikhil and {Schulze}, Steve and {Tanvir}, Nial R. and {Ackley}, Kendall and {Anderson}, Gemma and {Brammer}, Gabriel B. and {Christensen}, Lise and {Dhillon}, Vikram S. and {Evans}, Phil A. and {Fausnaugh}, Michael and {Fong}, Wen-fai and {Fruchter}, Andrew S. and {Fryer}, Chris and {Fynbo}, Johan P.~U. and {Gaspari}, Nicola and {Heintz}, Kasper E. and {Hjorth}, Jens and {Kennea}, Jamie A. and {Kennedy}, Mark R. and {Laskar}, Tanmoy and {Leloudas}, Giorgos and {Mandel}, Ilya and {Martin-Carrillo}, Antonio and {Metzger}, Brian D. and {Nicholl}, Matt and {Nugent}, Anya and {Palmerio}, Jesse T. and {Pugliese}, Giovanna and {Rastinejad}, Jillian and {Rhodes}, Lauren and {Rossi}, Andrea and {Saccardi}, Andrea and {Smartt}, Stephen J. and {Stevance}, Heloise F. and {Tohuvavohu}, Aaron and {van der Horst}, Alexander and {Vergani}, Susanna D. and {Watson}, Darach and {Barclay}, Thomas and {Bhirombhakdi}, Kornpob and {Breedt}, Elm{\'e} and {Breeveld}, Alice A. and {Brown}, Alexander J. and {Campana}, Sergio and {Chrimes}, Ashley A. and {D'Avanzo}, Paolo and {D'Elia}, Valerio and {De Pasquale}, Massimiliano and {Dyer}, Martin J. and {Galloway}, Duncan K. and {Garbutt}, James A. and {Green}, Matthew J. and {Hartmann}, Dieter H. and {Jakobsson}, P{\'a}ll and {Kerry}, Paul and {Kouveliotou}, Chryssa and {Langeroodi}, Danial and {Le Floc'h}, Emeric and {Leung}, James K. and {Littlefair}, Stuart P. and {Munday}, James and {O'Brien}, Paul and {Parsons}, Steven G. and {Pelisoli}, Ingrid and {Sahman}, David I. and {Salvaterra}, Ruben and {Sbarufatti}, Boris and {Steeghs}, Danny and {Tagliaferri}, Gianpiero and {Th{\"o}ne}, Christina C. and {de Ugarte Postigo}, Antonio and {Kann}, David Alexander},
        title = "{Heavy-element production in a compact object merger observed by JWST}",
      journal = {\nat},
         year = 2024,
        month = feb,
       volume = {626},
       number = {8000},
        pages = {737-741},
          doi = {10.1038/s41586-023-06759-1},
archivePrefix = {arXiv},
       eprint = {2307.02098},
 primaryClass = {astro-ph.HE},
       adsurl = {https://ui.adsabs.harvard.edu/abs/2024Natur.626..737L}
}

@ARTICLE{Yang22,
       author = {{Yang}, Jun and {Ai}, Shunke and {Zhang}, Bin-Bin and {Zhang}, Bing and {Liu}, Zi-Ke and {Wang}, Xiangyu Ivy and {Yang}, Yu-Han and {Yin}, Yi-Han and {Li}, Ye and {L{\"u}}, Hou-Jun},
        title = "{A long-duration gamma-ray burst with a peculiar origin}",
      journal = {\nat},
         year = 2022,
        month = dec,
       volume = {612},
       number = {7939},
        pages = {232-235},
          doi = {10.1038/s41586-022-05403-8},
archivePrefix = {arXiv},
       eprint = {2204.12771},
 primaryClass = {astro-ph.HE},
       adsurl = {https://ui.adsabs.harvard.edu/abs/2022Natur.612..232Y}
}

\appendix

\section{Redshift Evolution of the Distribution}
\begin{figure*}[h!]
\includegraphics[width=0.45\linewidth, height=5cm]{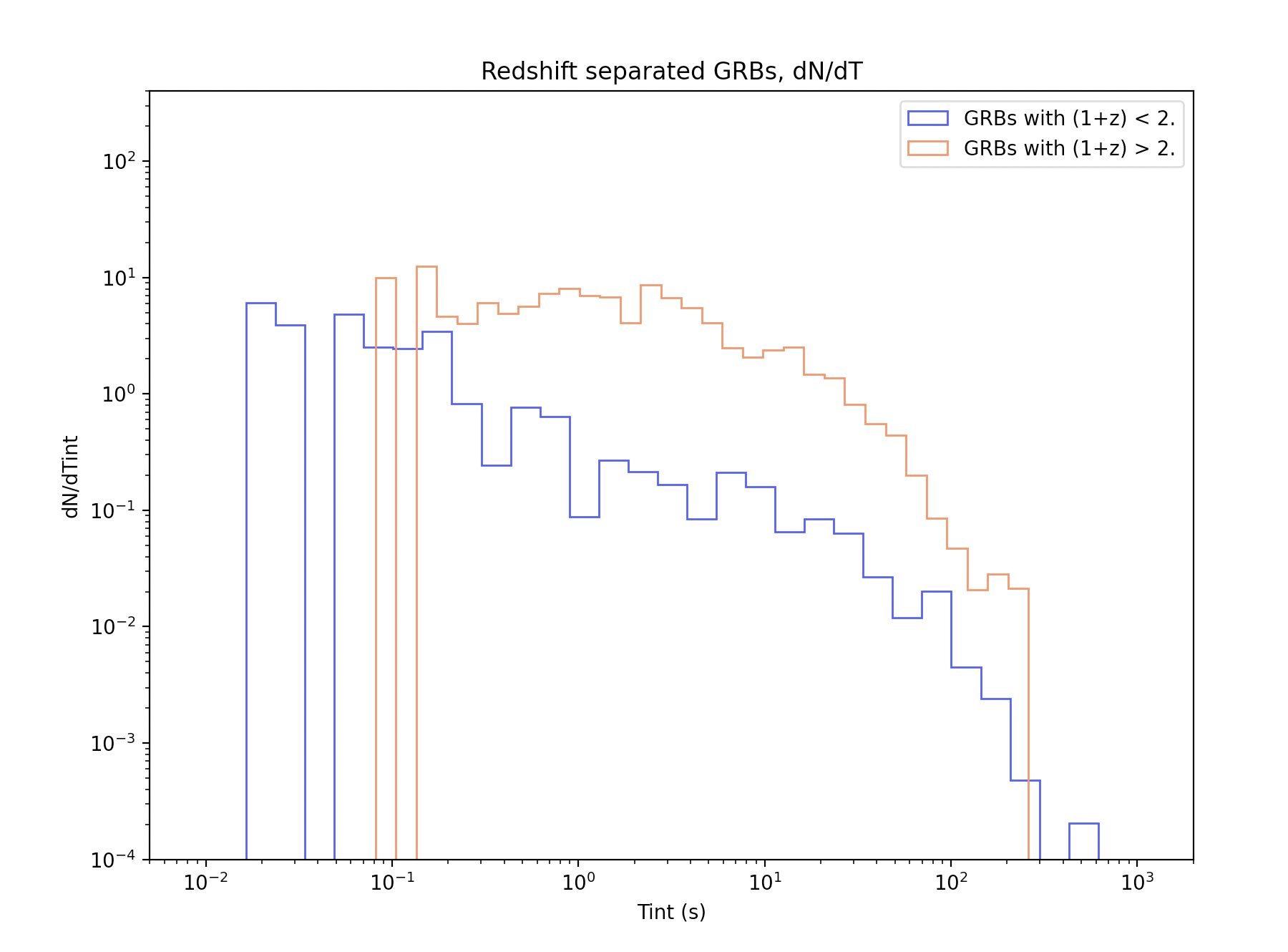}
\includegraphics[width=0.42\linewidth, height=5cm]{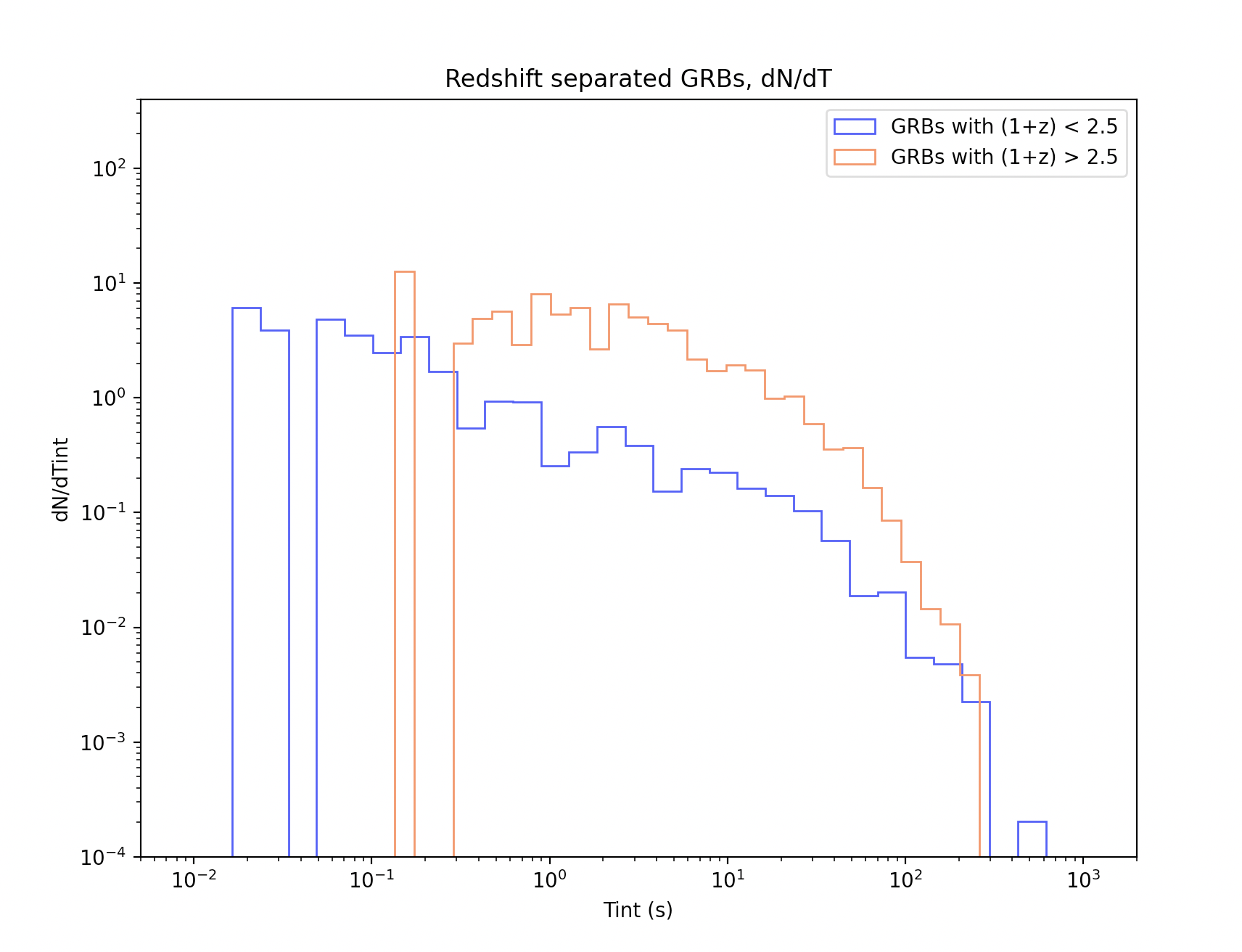}\\
\includegraphics[width=0.45\linewidth, height=5cm]{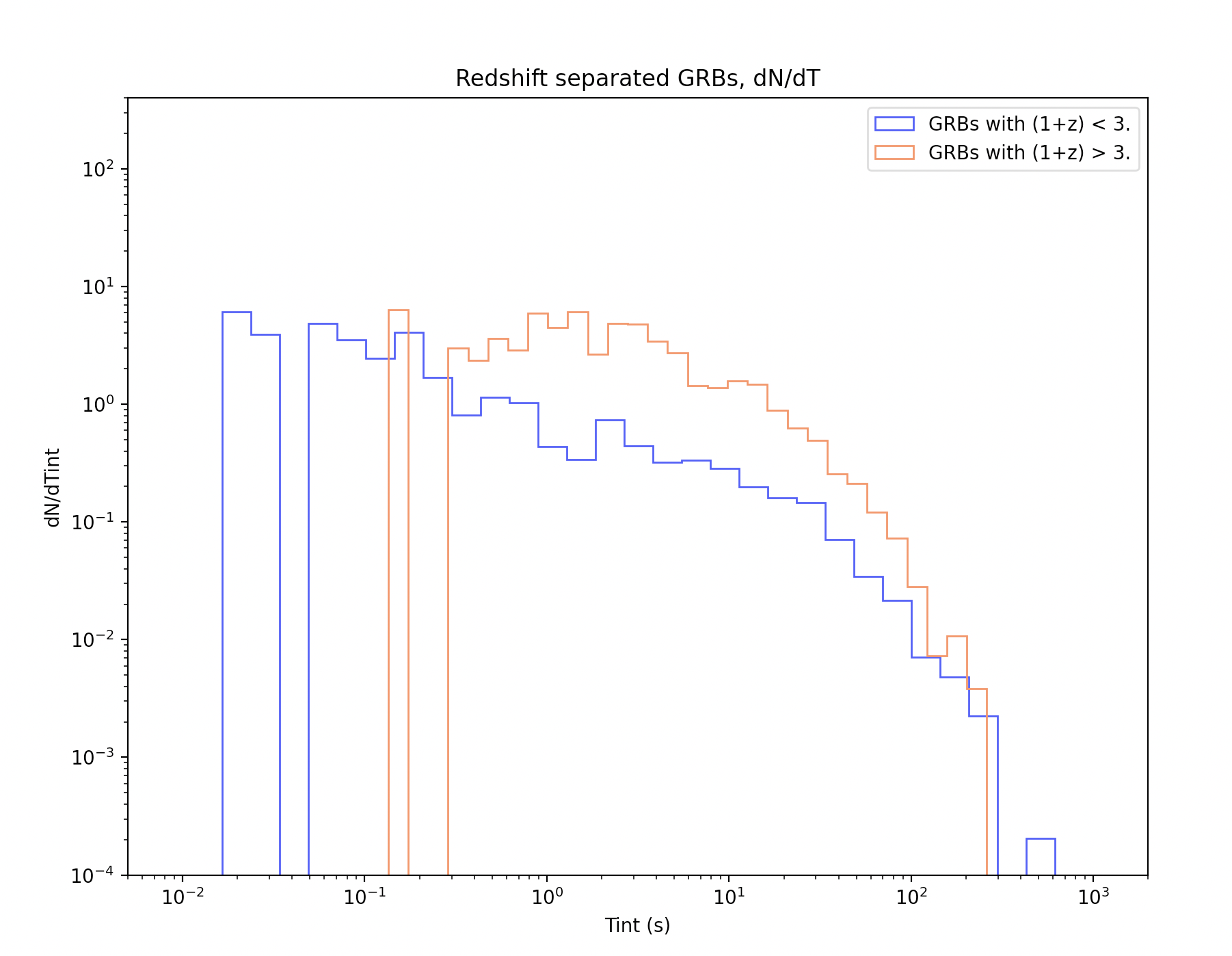}
\includegraphics[width=0.42\linewidth, height=5cm]{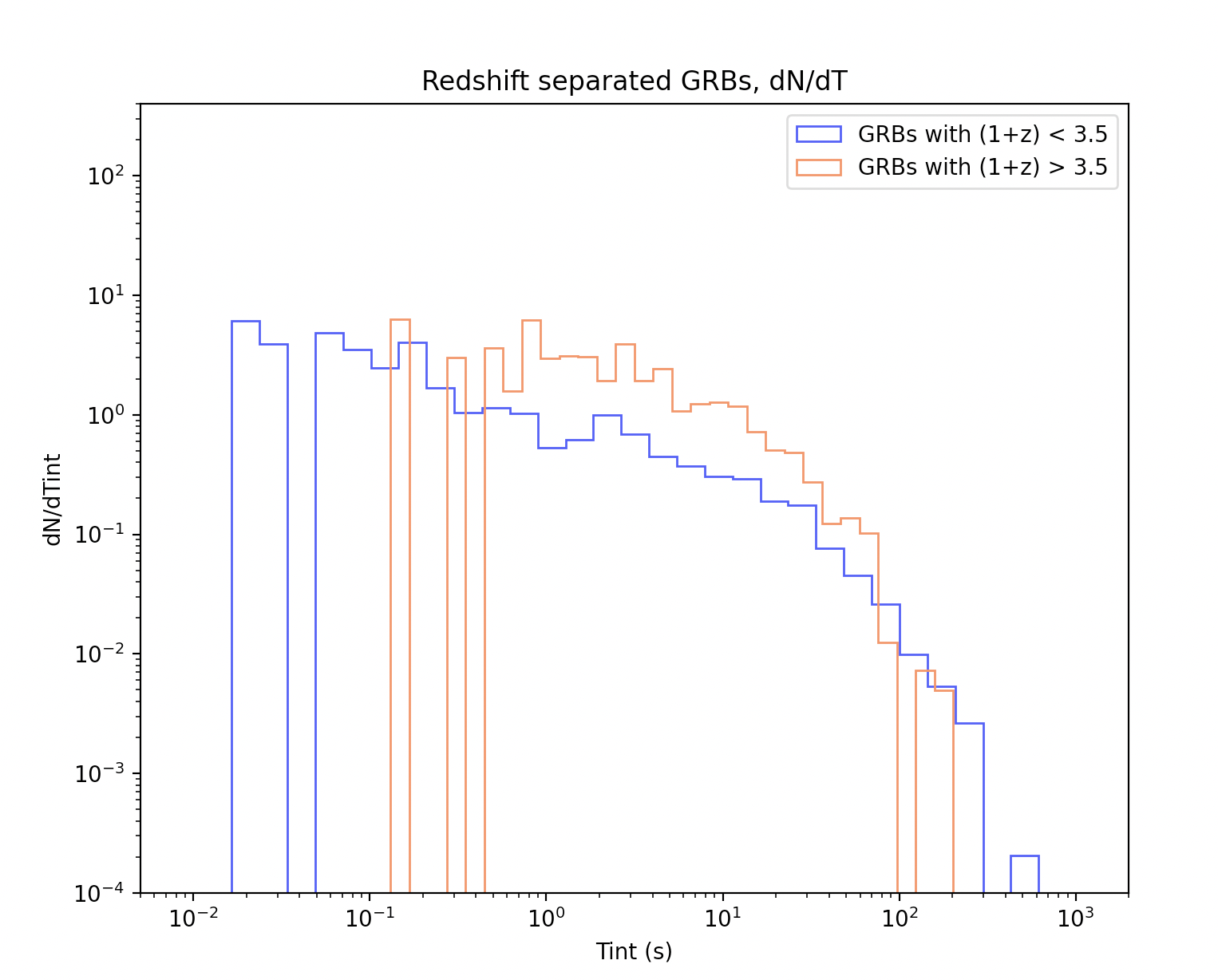}\\
\includegraphics[width=0.45\linewidth, height=5cm]{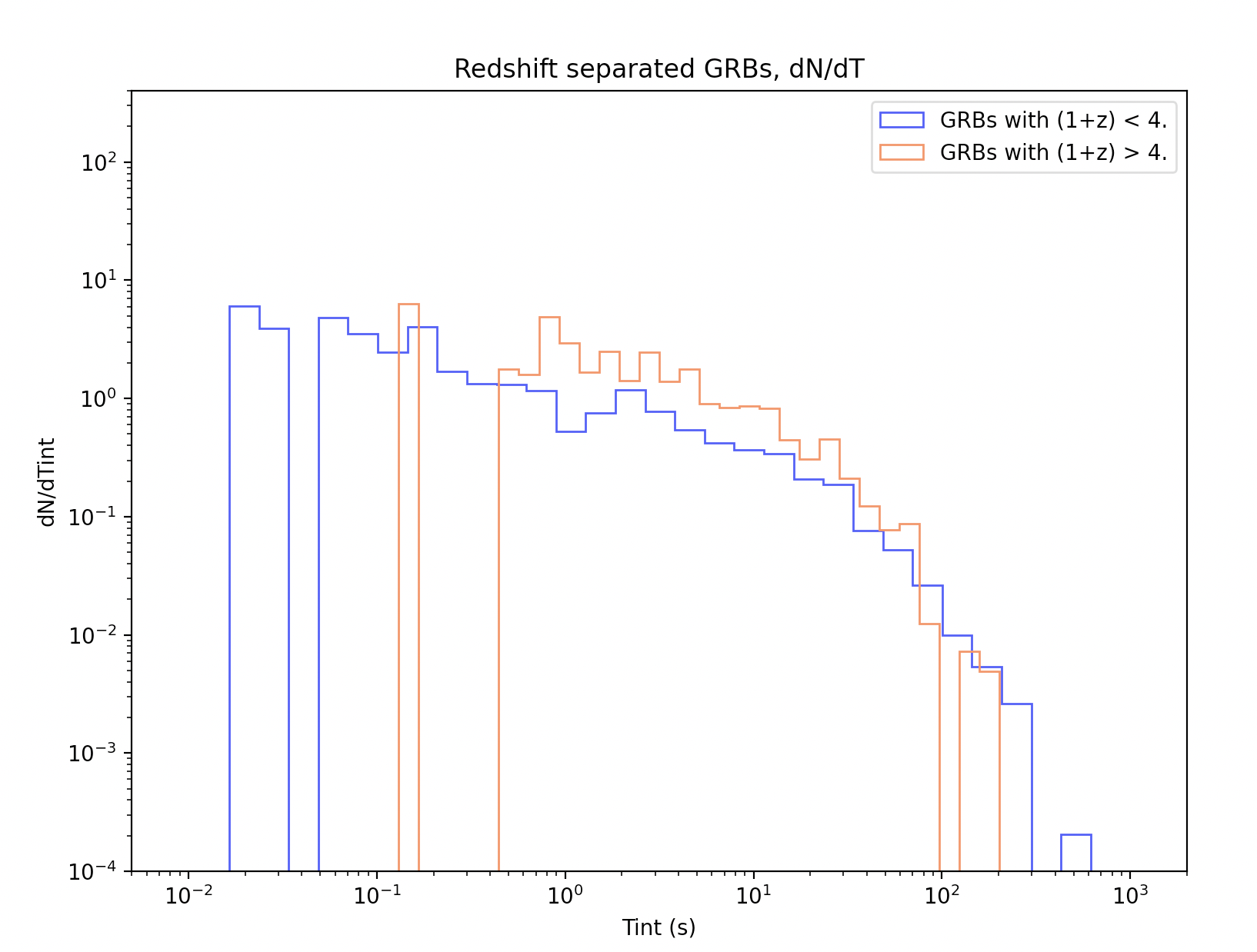}
\includegraphics[width=0.42\linewidth, height=5cm]{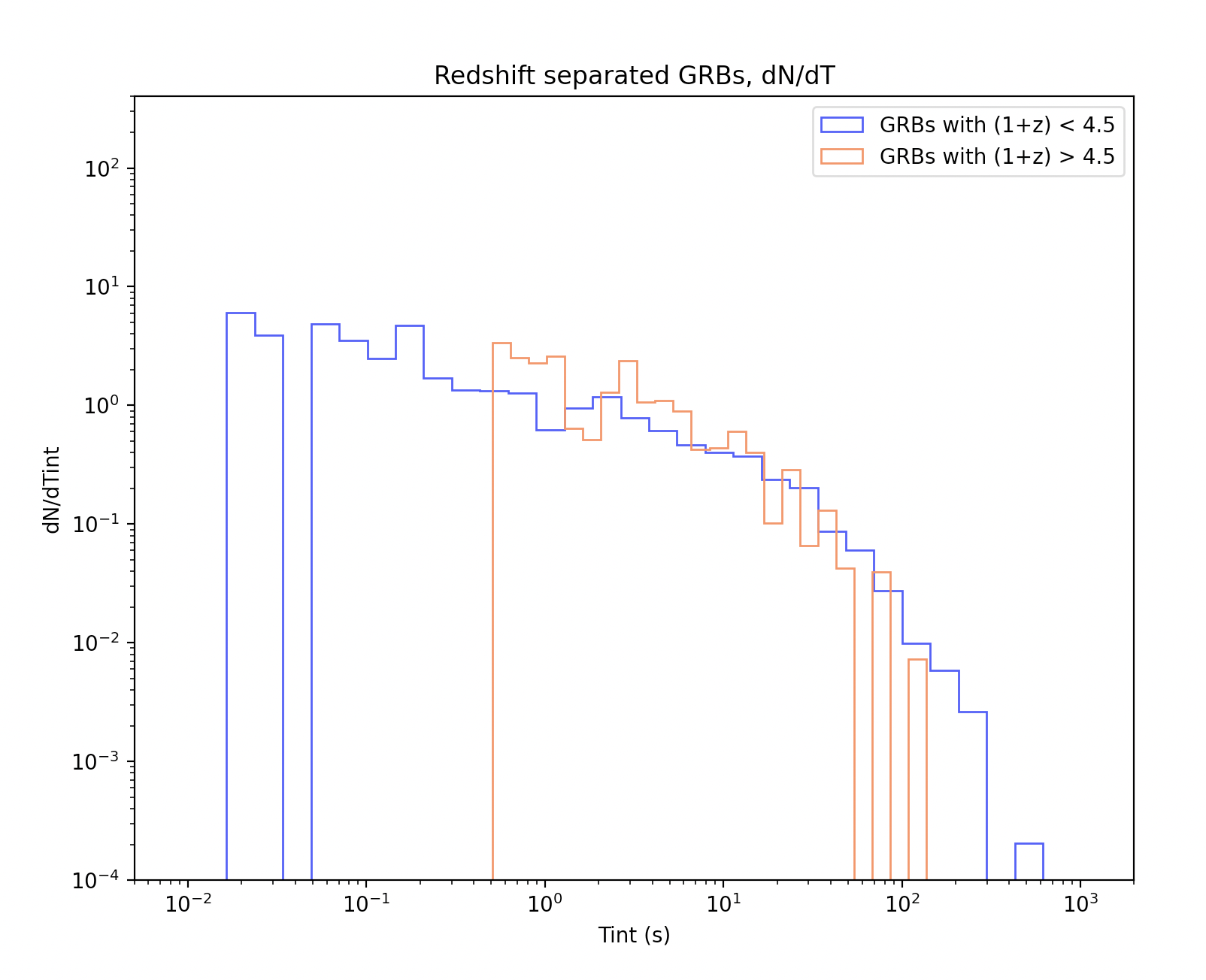}
\caption{Evolution of the redshift separated $dN/dT_{int}$ distribution for different values of the delimiting redshift, with "low" redshift are shown by blue lines and "high" redshift are showing by orange lines.}
%The sample sizes in each case are as follows, given in the following format: [Redshift separation $(1+z)$, number of low redshift bursts, number of high redshift bursts]. {\bf Top left:} [2., 165, 402], {\bf Top right:} [2.5, 261, 306], {\bf Middle Left:} [3.0, 331, 236], {\bf Middle Right:} [3.5, 406, 161], {\bf Lower left:} [4.0, 449, 118], {\bf Lower Right:} [4.5, 489, 78]. 
\label{fig:dNdTzevolve}
\end{figure*}

 Figure~\ref{fig:dNdTzevolve} shows the intrinsic duration distributions, separated into ``low'' redshift (blue lines) and ``high'' redshift (orange lines) sub-samples for different values of the redshift delimiter, ranging from $2 \lesssim (1+z) \lesssim 4.5$.  It is clear that separating the sample at a redshift that falls between $(1+z) \sim 2$ to $(1+z) \sim 3$ shows a clear difference in the $dN/dT$ distributions, in that a plateau is present in the high redshift sample, but not in the low redshift sample.  If sample biases and completeness are not playing too large of a role in these distributions, this would indicate non-collapsar progenitors begin to dominate the GRB distribution below redshifts of about $(1+z) \sim 2$, while there is a clear signature of the collapsar progenitor above this redshift. \\ \\

\section{Hard/Soft Division}
\begin{figure*}[h]
\includegraphics[width=0.5\linewidth, height=7.5cm]{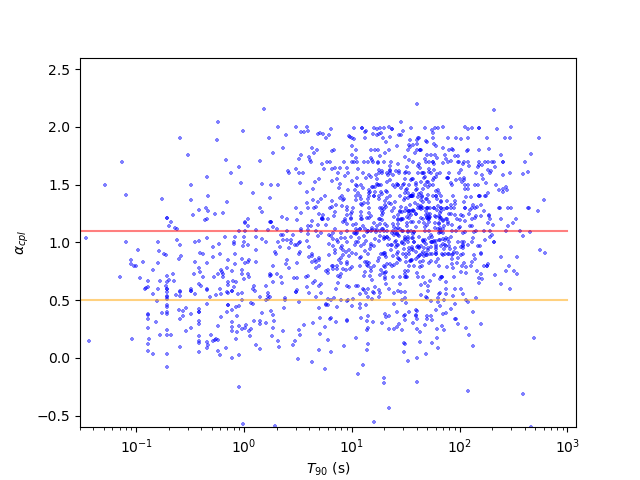}\includegraphics[width=0.5\linewidth, height=7.5cm]{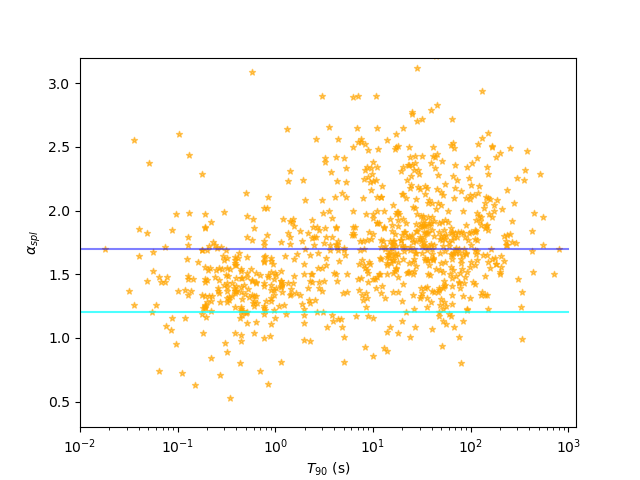}\\
\caption{Photon spectrum power law index vs $T_{90}$ for a cutoff power-law model (left panel) and a single power-law model (right panel). Above the upper lines defines our soft sample for each model; below the lower lines defines our hard sample.}
\label{fig:hardsep}
\end{figure*}

Figure ~\ref{fig:hardsep} shows the single power-law (right panel) and cutoff power-law (left panel) spectral indices, defined in equations 2 and 3 respectively, as a function of $T_{90}$ for {\em Swift} GRBs.  A higher $\alpha$ indicates a softer burst.  The traditional long-soft/short-hard separation is apparent.  Hence anything above the upper line in either figure is what we define as a ``soft'' burst.  Anything below the lower line in either figure is what we call a ``hard'' burst. \\

\section{Additional Progenitor and Jet Parameter Space}

%Whether the jet head is traveling relativistically or non-relativistically through the star depends on the luminosity of the jet $L_{j}$, the mass of the star $M_{*}$, the jet opening angle $\theta_{j}$, the radius of the star $R_{*}$, and the density profile of the star parameterized by $\rho(r) \propto r^{-\zeta}$. We expect the jet head to be traveling non-relativistically in the denser core of the star and potentially reach relativistic velocities as it traverses the outer parts of the star. 
Below, we provide additional contour plots of the threshold time calculated in section 4.3, for a broader range of jet and progenitor properties than shown in the main text in Figure~\ref{fig:Thresholdfull1}, where we used the average observed jet luminosity and opening angle values from our high redshift sample in our calculations.  Following our display in the main text, for each of the plots in Figure 8, the left panels show the calculation according to equation~\ref{eq:t_th}, while the right panels show this timescale with a numerical correction applied according to the simulations of \cite{Harr18}.\\

The top panels show the threshold time for a $40 M_{\odot}$ star (with all other variables as in the main text), the middle panels show the jet threshold time for a jet luminosity one tenth of the average observed value, and the bottom panels show the threshold time for a jet opening angle half of the average observed value.  As in the main text the black dotted and solid lines show a threshold time of 5 seconds and 15 seconds, respectively.\\

As expected, a larger progenitor mass will increase the threshold time, as will a lower luminosity jet.  Meanwhile, a smaller opening angle will decrease the time it takes the jet to propagate through the star and therefore requires larger progenitor radii to be consistent with the plateau end timescale.

\begin{figure}[h!]
\begin{centering}
\includegraphics[width=0.48\linewidth]{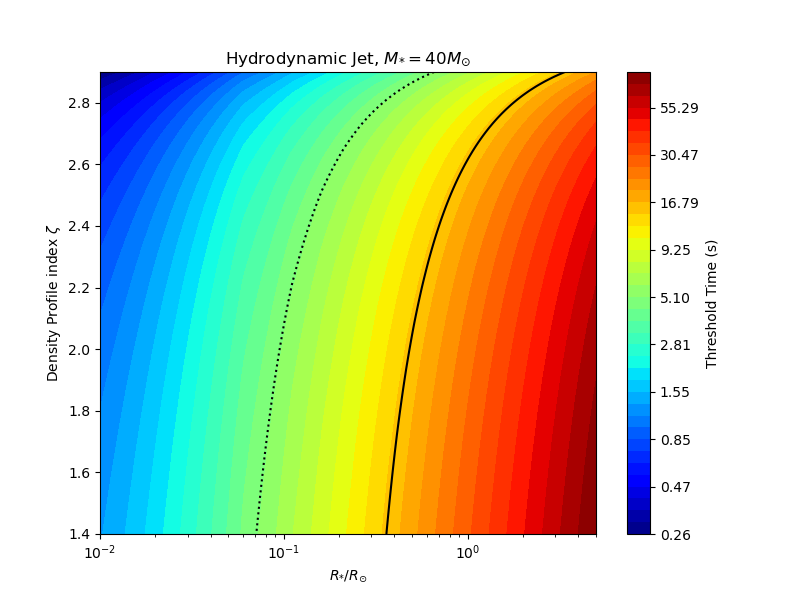}\includegraphics[width=0.48\linewidth]{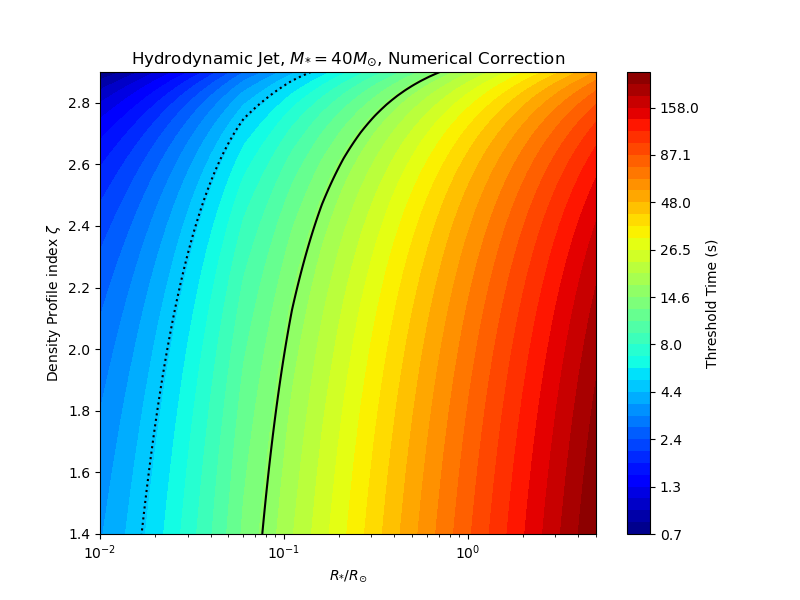}\\
\includegraphics[width=0.48\linewidth]{ 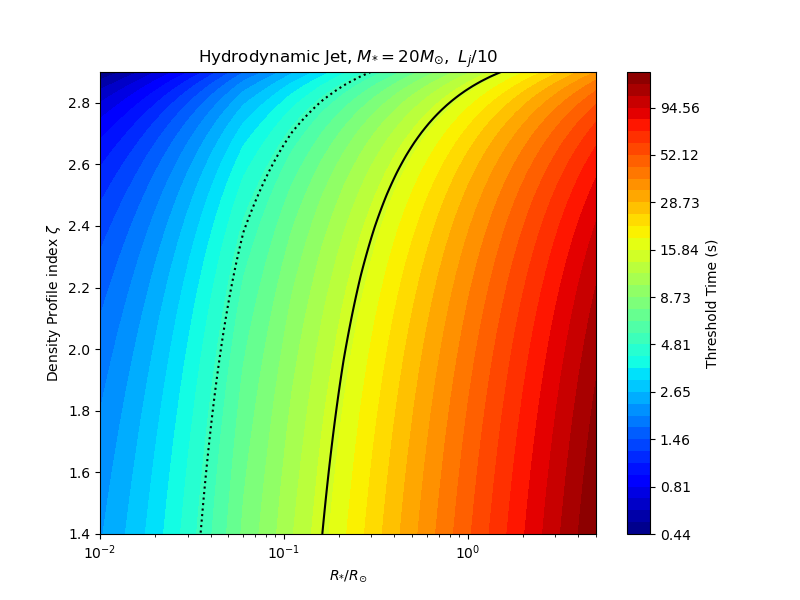}\includegraphics[width=0.48\linewidth]{ 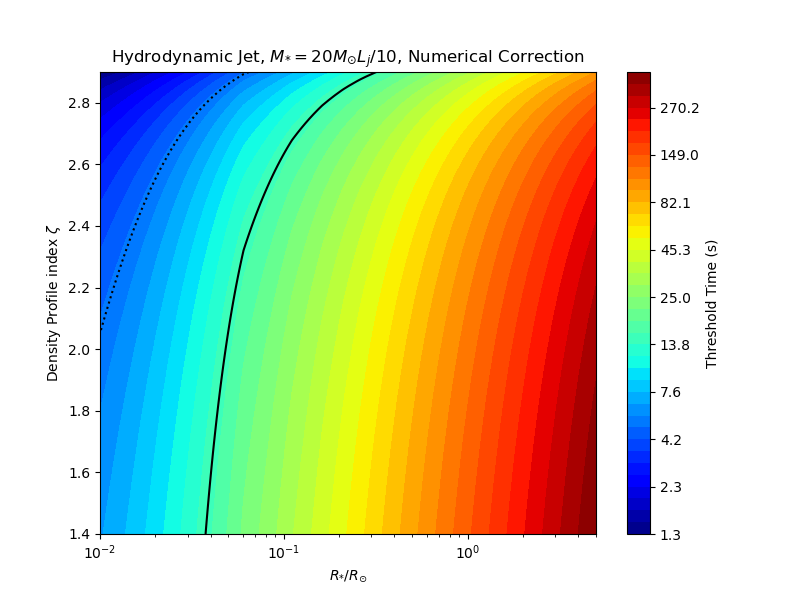}\\
\includegraphics[width=0.48\linewidth]{ 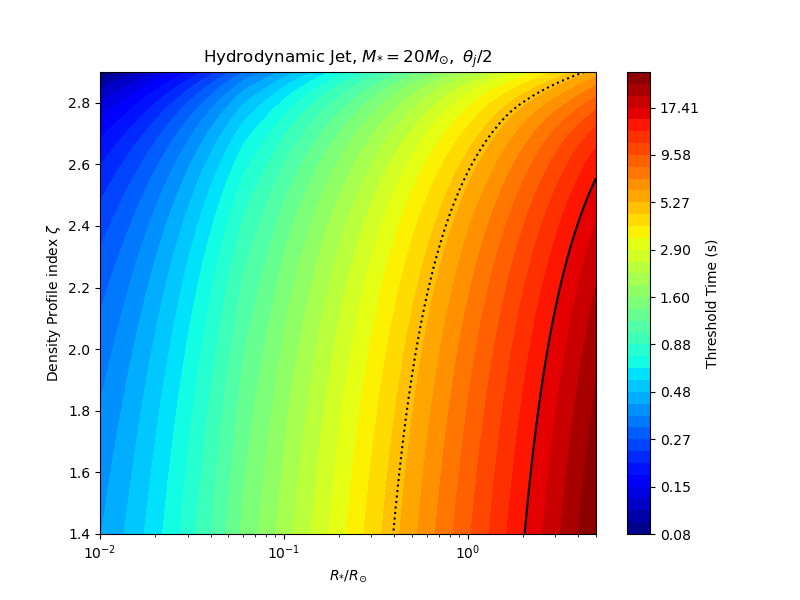}\includegraphics[width=0.48\linewidth]{ 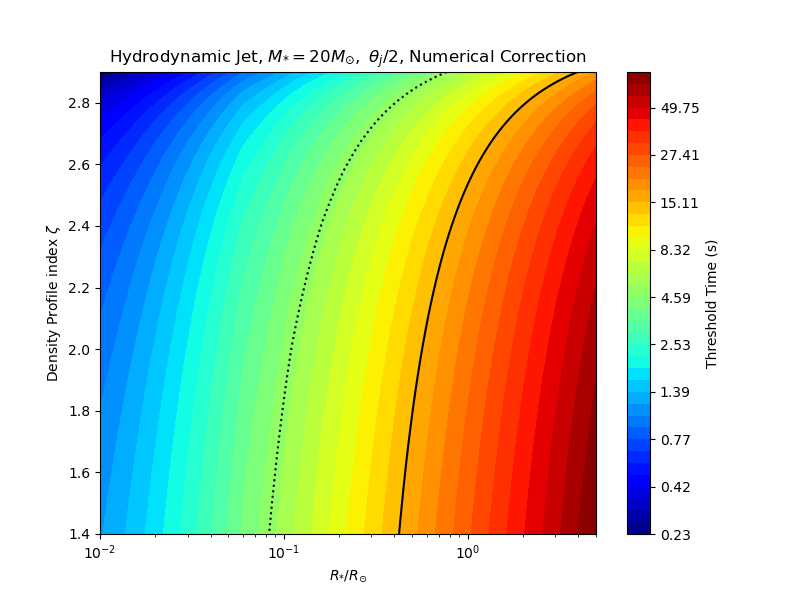}\\
\caption{Threshold time for different progenitor masses, jet luminosity and jet opening angle. the left panels show the calculation according to equation~\ref{eq:t_th}, while the right panels show this timescale with a numerical correction applied according to the simulations of \cite{Harr18}. The top panels show the threshold time for a $40 M_{\odot}$ star (with all other variables as in the main text), the middle panels show the threshold time for a jet luminosity $L_{j} = 2 \times 10^{49}$erg/s, one tenth of the average observed value of our high redshift sample, and the bottom panels show the threshold time for a jet opening angle $\theta_{o} = 2.9^{o}$ half of the average observed value of our high redshift sample.}
\end{centering}
\label{fig:AddContour}
\end{figure}

\end{document}